\documentclass{aa}  

\usepackage{graphicx}
\usepackage{txfonts}

\newcommand\mll{$M/L_{UV}$}
\newcommand\ml{$\rm M/L$}
\newcommand\muv{$M_{UV}$}
\newcommand\lgm{$\log (M/M_\odot)$}
\newcommand\ebv{E(B$-$V)}

\usepackage{enumitem}
\usepackage[pdfencoding=auto,psdextra]{hyperref}
\hypersetup{
    colorlinks=true,
    linkcolor=blue,
    filecolor=magenta,      
    urlcolor=blue,
    citecolor=blue
}
\begin{document} 

   \title{
   The early maturity of high-redshift galaxies:    Insights from sSFR, \ml, and SFHs at $z\sim 7-14$
   }

\titlerunning{The early maturity of high-$z$ galaxies}

   \author{P.~Santini \inst{1}          
        \and
        M.~Castellano \inst{1}
        \and
        A.~Calabr\`o \inst{1}
        \and
        A.~Fontana \inst{1}
        \and
        E.~Merlin \inst{1}
        \and
        D.~Bevacqua \inst{1}
        \and
        P.~Bergamini \inst{2}
        \and
        M.~Boquien \inst{3}
        \and
        S.~Cantarella \inst{4,5}
        \and
        L.~Ciesla \inst{6}
        \and
        A.~Ferrara \inst{7}
        \and
        S.~L.~Finkelstein \inst{8,9}
        \and
        F.~Fortuni \inst{1}
        \and
        G.~Gandolfi \inst{1}
        \and
        T.~Gasparetto \inst{1}
        \and
        E.~Giallongo \inst{1}
        \and
        N.~A.~Grogin \inst{10}
        \and
        S.~T.~Guida \inst{1,11,12}
        \and
        A.~M.~Koekemoer \inst{10} 
        \and
        N.~Menci \inst{1}
        \and
        L.~Napolitano \inst{1}
        \and
        D.~Paris \inst{1}
        \and
        L.~Pentericci \inst{1}
        \and
        B.~Perez-Diaz \inst{1}
        \and
        B.~Stoyanova \inst{1,13}
        \and
        T.~Treu \inst{14}
          }

   \institute{INAF - Osservatorio Astronomico di Roma, Via di Frascati 33, 00078 Monte Porzio Catone, Italy\\
              \email{paola.santini@inaf.it}
    \and
    INAF - Osservatorio di Astrofisica e Scienza dello Spazio di Bologna,
Via Piero Gobetti 93/3, 40129 Bologna, Italy
        \and
        Universit\'e C\^ote d’Azur, Observatoire de la C\^ote d’Azur, CNRS, Laboratoire Lagrange, F-06000 Nice, France
        \and        Astronomy Section, Department of Physics, University of Trieste, Via G.B. Tiepolo 11, 34143, Trieste, Italy
        \and
    INAF – Astronomical Observatory of Trieste, Via G.B. Tiepolo 11, 34143 Trieste, Italy
        \and
         Aix Marseille Univ, CNRS, CNES, LAM, Marseille, France
         \and
         Scuola Normale Superiore, Piazza dei Cavalieri 7, 50126 Pisa, Italy
    \and 
    Department of Astronomy, The University of Texas at Austin, Austin, TX, USA
    \and 
    Cosmic Frontier Center, The University of Texas at Austin, Austin, TX 78712, USA
         \and
    Space Telescope Science Institute, 3700 San Martin Drive, Baltimore, MD 21218, USA
    \and 
    Università di Napoli “Federico II”, C.U. Monte Sant’Angelo, Via
Cinthia, 80126 Napoli, Italy
    \and
    Max-Planck-Institut für extraterrestrische Physik, Gießenbachstraße 1, 85748 Garching, Germany
    \and 
    Physics Department, Tor Vergata University of Rome, Via della Ricerca Scientifica 1, 00133 Rome, Italy
    \and
    Department of Physics and Astronomy, University of California, Los Angeles, 430 Portola Plaza, Los Angeles, CA 90095, USA
             }

   \date{}

 
  \abstract 
{
The James Webb Space Telescope (JWST) has revealed an unexpected excess of UV-bright galaxies at $z>10$, unaccounted for by extrapolations from pre-JWST observations and theoretical models. Understanding the physical properties and star formation histories (SFHs) of  high-redshift systems is key to distinguishing between the different proposed scenarios and assessing their implications for the galaxy formation and evolution picture. We identified and analysed a sample of 2420 robust candidates at $z\sim 7-14$ drawn from the ASTRODEEP-JWST dataset over $\sim$0.2 deg$^2$, and modelled their properties with non-parametric SFHs to derive the specific star formation rate (sSFR) and stellar population properties. We find that the median sSFR and M/L remain roughly constant across the probed redshift range. We show that this result is robust against potential systematics unless a hidden population of dust-enshrouded starbursts, undetectable in the current data, exists at these redshifts. In any case, the absence of observed high-sSFR systems at the highest redshifts suggests that any dust-free starburst phase must be short-lived. The observed sSFR evolution is in tension with most (though not all) theoretical models, making it a key quantity for discriminating among competing scenarios. The sample shows a wide range of physical conditions and galaxy classes, including systems with low sSFRs and high mass-to-light ratios (M/L) up to $z\sim 10$, indicative of already-evolved galaxies only a few hundred megayears after the Big Bang, and different degrees of dust attenuation. We finally reconstructed the assembly histories of two sub-samples, namely the highest-M/L galaxies at $z\sim7-8$, which appear to have formed the bulk of their stars at least 500 Myr before observation, implying progenitors observable as UV-bright sources at $z>20$, and the full sample of $z>11$ galaxies, which formed through stochastic SFHs, remaining UV-faint for most of their early evolution, before undergoing recent ($\sim$50 Myr old) episodes of major growth.
}
%

   \maketitle
%

\section{Introduction}\label{sec:intro}

Since the start of science operations, 
the James Webb Space Telescope (JWST) has  
demonstrated its  revolutionary power for probing the early Universe, especially the first few hundred million years after the Big Bang,  
pushing the detection frontier to $z$$\sim$10-15 \citep[e.g.][]{castellano22,deugenio24,carniani24,napolitano25b,naidu26} and possibly beyond \citep{kokorev25a,perez-gonzalez25,castellano25,gandolfi25a,gandolfi26}. 
One of the most exciting and consolidated results 
is the evidence of an unexpectedly high number of rest-frame UV-bright $z \gtrsim 10$ galaxies \citep[e.g.][]{castellano22,castellano23,naidu22,harikane23a,donnan23,finkelstein23,finkelstein24,mcleod24,whitler25} confirmed by successful spectroscopic follow-up with NIRSpec \citep[e.g.][]{arrabal-haro23,curtis-lake23,harikane24,castellano24,carniani24,napolitano25b}. A very wide range of physical interpretations have been proposed to explain this excess. The first class of models invokes a brightening of the objects due to peculiar star formation (SF) properties, for example top-heavy initial mass functions \citep[IMF; e.g.][]{inayoshi22}, heavily stochastic SF \citep[e.g.][]{mason23}, and negligible dust attenuation \citep[e.g.][]{ferrara23,narayanan26}. Other models involve an accelerated evolution of galaxies  
due to enhanced SF efficiency and/or reduced feedback \citep[e.g.][]{dekel23,somerville25,cantarella26}, or even an accelerated formation due to potential modifications of the standard cosmological scenario \citep[e.g.][]{liu22,padmanabhan23,melia23,menci24}. The presence of active galactic nuclei (AGN) in high-z galaxies \cite[e.g.][]{harikane23b,bunker23,castellano24,napolitano25a} could also contribute to this excess \citep{hegde24}, although 
several sources responsible for the excess do not show any AGN feature \citep[e.g.][]{curtis-lake23,carniani24,napolitano25b}. The origins of this overabundance are therefore tied to fundamental properties of the high-redshift Universe, and motivate the great interest that this problem has raised recently.

Key to discriminating among the proposed scenarios is the investigation of galaxy physical properties and star formation histories (SFH). Thanks to JWST, the physical properties of a galaxy can be measured with significantly greater accuracy than  was possible before \citep[i.e. a factor of 5-10 lower relative uncertainties on the stellar mass,][]{santini23}. Furthermore, recent advances in modelling techniques have greatly improved our ability to reconstruct the  SFHs of galaxies 
\citep[e.g.][]{leja19,ciesla24,wang25,carvajal-bohorquez25}.

Another surprising result emerging from 
JWST observations is the remarkable diversity in the physical properties of high-redshift ($z>7$) galaxies. This variety is seen across multiple aspects of galaxy evolution: 
stellar population properties \citep[e.g.][]{santini23,conselice25,tang26}; 
SFHs \citep[e.g.][]{dressler24,endsley25}; 
quenching timescale, with passive galaxies found at very early epochs \citep[e.g.][]{russell25,weibel25,merlin25,baker25}; 
 physical conditions of the interstellar medium in terms of metal abundances, ionisation, and excitation conditions \citep[e.g.][]{schaerer22,cameron23,curti23,roberts-borsani26}; and 
structure and morphology  \citep[e.g.][]{kartaltepe23,treu23,lee24}

For this paper we focused our attention on the evolution of the specific star formation rate (sSFR), i.e. the SFR per unit stellar mass ($M$), and 
on the mass-to-light ratio (\mll, \ml~in the following), i.e. the ratio of the stellar mass to the rest-frame UV luminosity, which traces 
the nature of the stellar populations. By combining SF and stellar population properties, 
we tried to reconstruct the assembly history of our galaxies. 
To achieve this goal we exploited the ASTRODEEP-JWST database \citep{merlin24}, consisting of more than 500 thousand 
sources over an area of 0.2 deg$^2$. 
Assembling a statistically significant galaxy sample is essential in order to get rid of the large uncertainties that may affect individual fits \citep{wang25}.

This paper is organised as follows. 
We describe the dataset and sample selection in Sect.~\ref{sec:data}. In Sect.~\ref{sec:methods} we illustrate the methodology and discuss possible biases and completeness issues. 
We present our results in Sect.~\ref{sec:results} and discuss possible interpretations in Sect.~\ref{sec:disc}. Finally, we summarise our findings in Sect.~\ref{sec:summary}.  
We adopt the Planck18 \citep{Planck2018} Lambda cold dark matter ($\Lambda$CDM)  
cosmology and a \cite{chabrier03} IMF. Magnitudes are given in the AB system \citep{oke_gunn83}.\\\\
\section{Dataset}\label{sec:data}

\subsection{The ASTRODEEP-JWST catalogues}

This work is based on the extensive ASTRODEEP-JWST database publicly released\footnote{\url{http://www.astrodeep.eu/astrodeep-jwst-catalogs/}} by our group \citep{merlin24}. It consists of a set of multiwavelength photometric catalogues extracted from eight JWST programs targeting seven distinct areas in the sky. For the ABELL2744 fields we  used data from GLASS-JWST \citep{treu22}, UNCOVER \citep{bezanson24}, DDT 2756 (P.I. Chen) and GO 3990 (P.I. Morishita); for EGS we exploited the CEERS survey \citep{finkelstein25}; JADES \citep{eisenstein26} data were used for the GOODS-S and GOODS-N fields; a fraction of the Extended CDFS was mapped by the NGDEEP program \citep{bagley24}; finally, the COSMOS and UDS fields were observed as part of the PRIMER program (GO 1837, P.I. Dunlop).

Photometric catalogues are based on eight JWST NIRCam bands (F090W, F115W, F150W, F200W, F277W, F356W, F410M, and F444W)
and eight HST bands (ACS F435W, F606W, F775W and F814W, and WFC3 F105W, F125W, F140W, and F160W), although not all fields are provided with the full 16-band photometry. 
Source detection was performed 
on a weighted stacked F356W and F444W  image, and aperture photometry was measured with the \texttt{A-PHOT} software \citep{merlin19b} on PSF-matched images.  
In total, the ASTRODEEP-JWST database contains  $\sim530$k sources and covers a sky area of $\sim0.2$ deg$^2$.

For this work, we used the photometric redshifts 
estimated with the \texttt{zphot} software package \citep{fontana00} and released with the ASTRODEEP-JWST database, as well as the associated probability distribution functions $P(z)$. They were obtained by assuming \cite{bc03}
 stellar templates, exponentially declining star formation histories, and including nebular emission according to \cite{castellano14} and \cite{schaerer09}. 
 When available, we adopted spectroscopic redshifts instead of photometric ones. 
The ASTRODEEP-JWST release also includes $\sim$21000 high-quality spectroscopic redshifts available in the literature at the time of release, against which photometric redshifts were tested. 
We complemented these with 
6031 new high-quality spectroscopic redshifts 
that were released in the meantime. They include sources {\it a)} taken from published works considering the collection of \cite{napolitano25b} and  \cite{napolitano26} (see references therein),  {\it b)} measured from the 
NIRSpec GO program 3073 \citep{castellano26} and {\it c)} obtained by querying the DAWN JWST Archive (DJA) archive \citep{heintz24,degraaff25}, limiting to robust redshifts validated from visual inspection (grade$=$3). 
Overall, the photometric redshift accuracy is very good and in line with the standards in the literature, yielding a median $dz=|z_{phot}-z_{spec}|/(1+z_{phot})$ of 0.006, NMAD\footnote{NMAD=1.48$\times$median(dz).} equal to 0.036 and an outlier fraction of 9\%, where outliers are defined as sources for which $dz>0.15$.  
We refer to \cite{merlin24} for 
detailed information on photometry and on photometric and spectroscopic redshifts.

 \subsection{Sample selection}\label{sec:sample}

To select a robust sample of high-redshift ($z\gtrsim7$) galaxies, 
we applied a combination of photometry and photo-$z$ criteria, starting from and customising the selection technique presented by \cite{finkelstein24} (see also \citealt{conselice25}). 
We tuned each  of the thresholds in the various steps of the selection  (required number of bands and associated minimum and maximum S/N, photometric redshift probabilities and best-fit $\chi^2$) by checking the performance against spectroscopic sources, aiming at the largest level of completeness keeping a low fraction of low-redshift interlopers.

\begin{enumerate}
\renewcommand{\labelenumi}{\roman{enumi})}

\item We started by rejecting objects with photometric issues, exploiting the \texttt{flag} columns in the ASTRODEEP-JWST catalogues. In particular, we rejected spurious detections   (including visually identified image defects), saturated or boundary sources and  bad measurements in both the F356W and F444W bands. We also excluded sources missing more than 4 JWST bands, to ensure robust sampling of the spectrum. These cuts reduce the catalogues to $\sim$90\% of their parent samples.

\item
We then cut the samples at S/N$>$5 in the F444W band, corresponding to $\sim 27.9-30.3$ mag, depending on the field ($\sim$62\% of the initial dataset). 

\item
We considered all sources with 
redshift higher than 6.5, i.e.  
$\sim$2.5\% of the initial sample, with the exception of COSMOS where the redshift cut selects almost 18\% of the sample due to a higher fraction of photo-$z$ failures,  mostly associated with image defects not properly masked by the \texttt{flag} catalogue column. 

\item
To deal with potential inaccuracies 
in the photometric redshift estimates and limit the analysis to robust high-redshift candidate galaxies, 
we required detection redwards of the Ly$\alpha$ break and non-detection bluewards of it. 
In particular, we required {\it a)} 
at least three detections redder than F090W/F115W/F150W/F200W, depending on the redshift, of which at least two with S/N$>$5,
and {\it b)}  at least one measurement shortwards of the break and none with S/N$\geq$2, considering ACS bands and the three bluest NIRCam bands, 
again depending on redshift. 
This photometric selection results in a further reduction of the samples to 
1.5\% of the initial number 
($\sim$59\% of the $z>6.5$ subsample with reliable photometry, and only 11\% for the COSMOS field, 
compensating the suspiciously large fraction of high-redshift sources).

\item 
We then applied selections based on the quality of the photometric redshifts, excluding all sources for which the reduced $\chi^2$ of the associated best-fit solution is larger than 10. We  
required that 70\% of the probability lies at $z>5.5$ and 
we rejected any source with a secondary redshift solution at $z<5.5$ whose associated probability is larger than 40\% of the best-fit probability.  This step of the selection further reduces the samples to 0.5\% of the parent catalogues ($\sim$20\% of the $z>6.5$ subsample with reliable photometry, and  3\% for COSMOS).
\end{enumerate}

Criteria iv) and v) are meant to clean the sample from lower-redshift interlopers and ensure robust redshifts. We therefore 
kept in the sample 
275 spectroscopic sources with redshift larger than 6.5 even if not meeting these two criteria 
due to lower S/N 
photometry. These spectroscopic galaxies allow us to extend the redshift range beyond $z\sim 12.5$. We checked that their inclusion, despite increasing the typical uncertainty at $z\gtrsim 8-9$, does not affect the main results of the paper. 

To evaluate the completeness and purity of our sample, we repeated the selections above on the  full spectroscopic subsample ignoring the spectroscopic redshifts and only relying on photometric ones. 
In the end, the completeness turns out to be 61\% on average (52-76\%, depending on the field, its photometric coverage and depth). Roughly half of the low-redshift sources incorrectly included in the selection are only slightly below the threshold, not affecting our results appreciably. The fraction of severe interlopers with $z_{spec}<$ 5.5 
is 4\%.  
To test whether these objects may affect our results, we ran the full analysis described in the following on all known interlopers (with spectroscopic redshift lower than 6.5) which, on the basis of photometric data alone, would be selected by our technique. These sources do not occupy any peculiar region of the diagrams presented in the following, implying that a small fraction of low-redshift sources erroneously included in the sample does not impact our conclusions.

Two final steps were performed in the selection. We visually inspected each candidate and excluded AGN-contaminated and stellar sources.

\begin{enumerate}
\renewcommand{\labelenumi}{\roman{enumi})}
\setcounter{enumi}{5}

\item
We visually inspected all sources and discarded problematic ones not flagged in the parent catalogue, i.e. defects, spikes, objects whose photometry is clearly affected by noise at the borders of the images or blending issues, and sources with poor fits. Visual inspection removes from a few to up to 11\% 
of  objects, depending on the field. 

\item
We removed:  
{\it {a)}} known AGN flagged as such in the ASTRODEEP-JWST database \citep{merlin24} or 
according to the recent literature (see references in \citealt{napolitano25a});  
{\it {b)}} Little Red Dots, cross-correlating our sample with the lists of \cite{kokorev24}, \cite{barro24}, \cite{perez-gonzalez24}, \cite{kocevski25}, and \cite{labbe25}; 
{\it {c)}} stars, using the SExtractor \texttt{CLASS\_STAR} classifier, which is reliable for bright sources, requiring $\texttt{CLASS\_STAR}>0.95$ \& S/N$>$30 in detection; {\it {d)}} 
known Brown Dwarfs by considering the lists of \cite{hainline24} and \cite{holwerda24}. Together, these classes of objects represent 5\% of the selected sources.

\end{enumerate}

\begin{figure}[t!]
    \centering
\includegraphics[width=9.5cm]{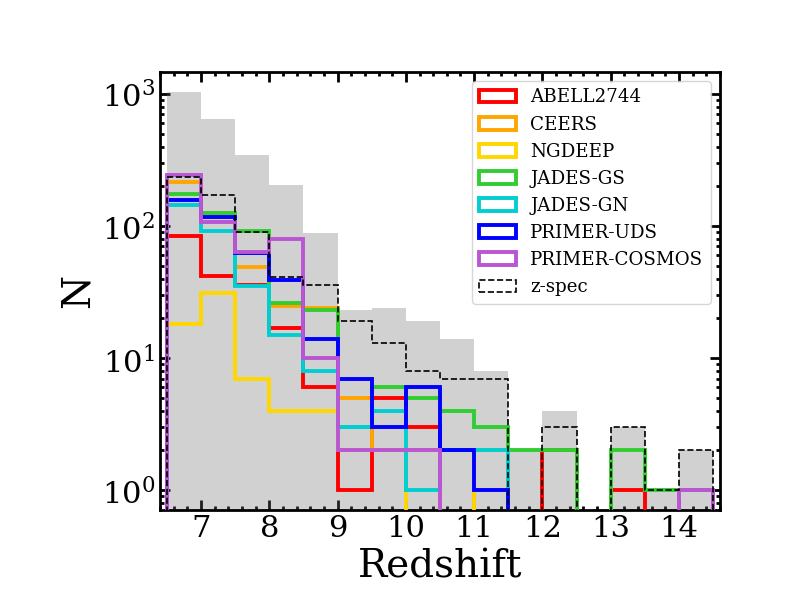}
\caption{Redshift distribution of the total sample (grey filled histogram) and of the galaxies selected in the different fields (coloured open histograms; see legend). The black dashed open histogram shows the distribution of spectroscopic sources.
}
\label{fig:zdist}
\end{figure}

In total, our sample includes 
 2418 (53) sources with redshift larger than 6.5 (10), of which 
 640 (31) are spectroscopic. 
The redshift distribution in the different fields is shown in Fig.~\ref{fig:zdist}.

\section{Methods}\label{sec:methods}
\subsection{Estimate of physical parameters}\label{sec:fit}

\begin{figure}[t!]
    \centering
\includegraphics[width=9.5cm]
{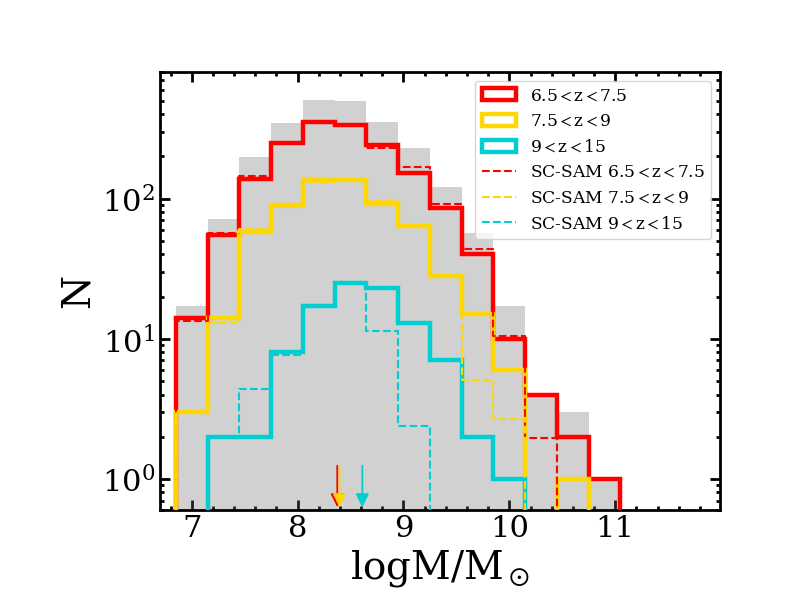}
\caption{Stellar mass distribution for the full sample (filled grey) and in three redshift bins (thick solid lines). The small coloured arrows indicate the median of the distribution in each redshift bin. The thin dashed lines show the mass distribution of the Santa Cruz simulation, mass-matched  to the observed galaxies (see Sect.~\ref{sec:massmatching}),  in the three redshift bins normalised to the peak of the observed mass distributions. 
}
\label{fig:Mdist}
\end{figure}

We estimated physical parameters with the CIGALE  SED fitting code \citep{boquien19}. 
We fitted the multiwavelength photometry at fixed redshift and added a 10\% relative error in quadrature to flux uncertainties to account for calibration and aperture corrections systematics \citep{harvey25}.

Growing evidence suggests that high-redshift galaxies have highly stochastic star formation histories (SFH), 
both based on theoretical arguments \citep[e.g.][]{muratov15,faucher-giguere18,furlanetto22} and on observational results \citep[e.g.][]{
ciesla24,cole25,endsley25,mcclymont25,kokorev25,carvajal-bohorquez25,tang26}. 
For this reason, we modelled our galaxies with flexible non-parametric SFHs, adopting the prescription proposed by
\citet{ciesla24}, shown by the authors to be well suited to high-redshift galaxies: the SFH is approximated by a given number of equally, linearly spaced  time bins of constant SFR with flat prior, which allows for a high level of stochasticity. 
The age of the oldest stellar populations 
can assume nine  equally spaced from 100 
to 900 Myr, never exceeding the age of the Universe at the galaxy redshift.  
Following 
\cite{ciesla24}, 
we adopted eight time bins plus a most recent 
bin of fixed duration $t_0$. By definition,  $t_0$ is the timescale over which the SFR is averaged. 
We adopted $t_0=20$ Myr to recover recent short-term SFH variations while avoiding excessive degeneracy, given that photometric data cannot reliably constrain variations of the SED on shorter timescales.
In fact, the fit turns out to be highly degenerate when fixing $t_0$ to a shorter duration, 
resulting in extremely large error bars on the SFR.  

We adopted the stellar population templates of \cite{bc03}, allowing stellar metallicities of 2\%, 20\% and 100\% Solar. Nebular emission was computed using the templates of \cite{inoue11} (see \citealt{boquien19} for details). We set the ionisation parameter to $\log U=-3$, $-2$ or $-1$, and allowed gas metallicities of 10\%, 20\% and 100\% Solar. 
Dust attenuation was modelled with the \cite{calzetti00} law ($R_V=4.05$), with colour excess on the lines \ebv$_{\rm line}$ spanning 0.15–1.7 (9 steps). The colour excess on the continuum, \ebv, was set to 0.44 times the value on the lines.

Since the physical parameters inferred from the SED fitting analysis might depend on the assumptions, we extensively tested that the overall results of the paper are robust against systematics. 
We describe in Appendix~\ref{app:fitsystematics} the different fitting setups considered as well as the results obtained. 

Physical parameters were inferred as the median of the probability distribution functions (PDF), 
with uncertainties given by the 
16${th}$--84${th}$ percentiles. 
 The sSFR and \ml~ratios, where L is rest-frame luminosity in the GALEX FUV filter (centred at 1500\AA) inferred from the fit, were derived by constructing the PDFs of the corresponding quantities across all models, weighted by their probabilities; again, the median  was adopted as the best estimate, and the 16${th}$--84${th}$ percentile range as the uncertainty. 
Uncertainties on the photometric redshifts were not considered. When accounting for them, 
we find only $\sim$2\% of sources, mostly below $z\sim 8.5$, having a (50\%) lower sSFR, and a relative error distribution slightly skewed towards higher values, mainly due to a tail of highly degenerate objects. Overall, the results of the paper remain unchanged, implying that photometric redshifts do not dominate the error budget. 

The ABELL2744 field is a 
lensed region centred on the galaxy cluster of the same name, 
and we inferred the intrinsic properties by correcting for magnification using the model of \cite{bergamini23}. The magnification factors range from $\sim 1$ to $\sim 35$, with a median  
value of $\sim 2$, 
allowing us to extend the analysis to lower stellar masses than would be possible in the absence of lensing. We note, however, that sSFRs and \ml~ratios, which are the core of this work, are unaffected by lensing. 

The mass distribution of our galaxy sample is shown in Fig.~\ref{fig:Mdist}. 
Massive galaxies  progressively disappear as redshift increases. 
The median value is \lgm~$\sim 8.4$ at z<9 and $\sim8.6$ at higher redshift, likely 
as consequence of incompleteness at low masses (see below and Appendix~\ref{app:completeness}) becoming more pronounced at high redshift.

We verified the accuracy of the inferred physical parameters in two ways. We first tested that the SFR agree with those based on spectroscopy, for the spectroscopic sub-sample. 
We then checked that the lack of optical constraints for the highest-redshift sources does not affect our results, particularly the stellar mass estimates.
To this aim, we fitted our sources ignoring the two reddest bands (F410M and F444W). This way, the resulting spectral coverage at $z\sim8-11$ is very similar to that achieved at higher redshifts when the full photometry is used.
Despite an increase in the uncertainties, the stellar masses turned out to be $\sim0.1$ dex larger on average, producing a similar offset on the sSFR and the \ml~at the highest redshifts.  This offset is however smaller than the typical statistical uncertainty and well within the dispersion of the sample. 
A detailed description of these consistency tests can be found in Appendix~\ref{app:accuracy}.

\subsection{Selection window and completeness} \label{sec:completeness}

Given the complex criteria adopted to select a robust sample of $z\gtrsim7$ galaxies and to assess potential biases, we exploit mock galaxies to explore the completeness of our sample at different masses and luminosities. Through this approach, we also assess our ability to recover galaxy properties.

We considered the mock catalogues extracted from the Santa Cruz semi-analytical model \citep[SC-SAM hereafter,][]{somerville15,somerville21}, and we complemented it with highly obscured (\ebv$\geq$0.5), star-forming  
galaxies simulated from stellar population synthesis modelling. We perturbed the simulated photometry in order to reproduce the typical photometric uncertainties of our galaxies,  we calculated the photometric redshifts and the physical parameters, and we applied the very same selection criteria used on our observed galaxies.  
The details of the simulation are provided in Appendix~\ref{app:completeness}, and the results are shown in Fig.~\ref{fig:simu}. We briefly summarise them here. 

When standard galaxies are considered (i.e. from the SC-SAM), the only serious completeness concerns affect low-mass ($<$$10^8 M_\odot$) and faint ($M_{UV}$>$-18$) galaxies with SFR$<$1 (2)~$M_\odot$/yr at $z$$<$10.5 ($z$$>$10.5), as expected, 
with a completeness of 20-40\%. The rest of the sample is instead highly complete, with a level of 70-90\% especially in the  $10^8-10^9M_\odot$ mass range, i.e. the typical mass of our sources from which we derive the redshift evolution of the sSFR and \ml. The situation is different when heavily dust-obscured (\ebv$>$0.5, i.e. $A_V\sim2$) sources are examined. 
 While dust attenuations as high as $A_V \sim 0.8-1.2$ 
have recently been  measured in $z\sim10-12$ galaxies \citep{mitsuhashi25,donnan25,rodighiero26}, 
$A_V> 2$ sources at these redshifts have not been spectroscopically confirmed 
yet, despite photometric hints of their existence \citep{rodighiero23}. 
Should these dusty galaxies exist, we could not observe them with our data  due to their extremely faint fluxes. We would only be able to select 28\% 
of them at $z<10$, and  almost none  ($<$3\%) at higher redshift. Overall, with the exception of this hypothetical population of high-$z$ extremely dusty galaxies, our technique can properly recover galaxy physical parameters.  

 Overall, based on our simulations, we stress that our results on the evolution of the overall properties of the sample apply to galaxies with stellar mass $>$$10^8 M_\odot$, $M_{UV}$<$-18$, SFR$\gtrsim$1-2$M_\odot$/yr, and low to moderate dust attenuation.

\subsection{Mass-matched model galaxies} \label{sec:massmatching}

The SC-SAM (see Appendix~\ref{app:completeness} for details) was also adopted  as one of reference models in this analysis for comparing our galaxies, 
as, unlike  others, it provides the full population scatter in addition to average values.  
Since the model is  typically characterised by lower-mass galaxies than the data sample, 
for an unbiased comparison we extracted a subset of model galaxies mass-matched to our observations.  To this aim, we defined a maximum tolerance in redshift and in mass of $\Delta z=0.1$ and $\Delta \log M=0.1$; for each observed galaxy, we picked 10 random simulated galaxies with consistent redshift and stellar mass, avoiding repetitions. For $\sim$13\% (7\%) of the galaxies, not enough matches were found, essentially due to the lack of simulated massive galaxies at the highest redshifts, so we lowered the redshift and  $\log M$ tolerance to 0.2 (0.3). In the end, only $\sim$2\% of the galaxies have no match, basically because the model does not extend to masses larger than $10^{10}M_\odot$. 

The resulting mass distribution of model galaxies used in our comparisons is shown in Fig.~\ref{fig:Mdist}.

\section{Results} \label{sec:results}

\begin{figure*}[t!]
    \centering
\includegraphics[width=\textwidth]{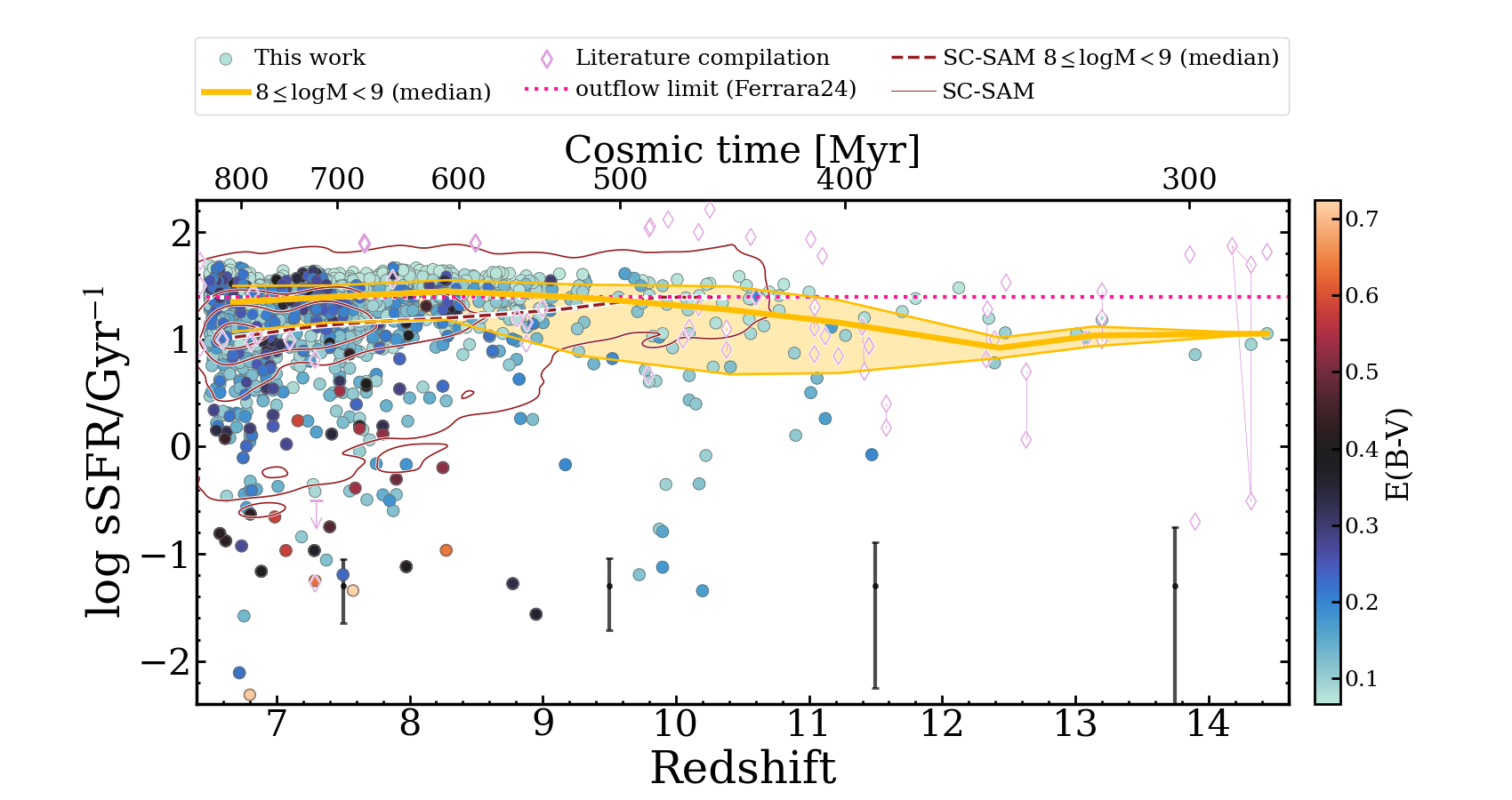}
\caption{Evolution of the sSFR. The median error bar in bins of redshift is shown at the bottom. The yellow line and shaded region represent the median and 16${th}$--84${th}$  percentile range of the \lgm~$=8-9$ 
subsample. 
Points are colour-coded according to the dust attenuation \ebv. The \ebv~was calculated on a discrete grid, and so for visualisation purposes we adopt the likelihood-weighted average. 
The pink open  
diamonds show a compilation of spectroscopic, JWST-based results taken from the recent literature   
\citep{tacchella23,robertson23,roberts-borsani23,arrabal-haro23,wang23,zavala24,deugenio24,hsiao24,hainline24b,harikane24,harikane25,looser24,carniani24,carniani25,topping25,trussler25,alvarez-marquez25,helton25,wu25,kokorev25,donnan25,weibel25,naidu26,witstok26,harikane26,rodighiero26,alvarez-marquez26,marques-chaves26}, with lines connecting different estimates for the same sources. 
Where available, we considered the SFR averaged over a timescale that is as close as possible to our choice of 20 Myr, sometimes averaging  estimates over different timescales.  
The magenta 
horizontal dotted line represents the sSFR threshold for developing radiation driven outflows capable of expelling dust according to the model of \cite{ferrara24a} (see text). 
The dark red dashed line and solid thin  curves are the median and 10\%, 50\%, 80\%, and 99.9\% probability densities for the SC-SAM. 
}
\label{fig:ssfr}
\end{figure*}

\subsection{The evolution of the specific SFR}\label{sec:ssfr}

The redshift evolution of the sSFR is shown in Fig.~\ref{fig:ssfr}. While the median sSFR is known to increase from the local Universe to $z\sim6$ \citep[see][and references therein]{santini17}, our results indicate that it remains approximately constant over the previous $\sim$600 Myr of cosmic time.  
This is clear when looking at galaxies with similar mass. 
The yellow line shows the median trend for $10^8-10^9 M_\odot$ galaxies (a similar lack of evolution is also observed for lower- and higher-mass galaxies). 
In agreement with the median trend, the majority ($\sim$75\%) of sources have relatively high sSFR ($\sim 20-40$ Gyr$^{-1}$). 
However, we observe notable outliers exhibiting low sSFR up to $z\sim 10$, discussed in Sect.~\ref{sec:passive}.

Our measurements show general consistency with previous results, shown by the pink open diamonds, with the exception of the highest values. Part of the mismatch may be ascribed to different assumptions in the estimate of physical parameters, as also shown by the scatter in literature results obtained by different studies for the same sources (open symbols connected by lines in Fig.~\ref{fig:ssfr}). 
One key assumption is the SFH, including  
our choice of a duration of $t_0=20$ Myr for the most recent time bin. Averaging over such a timescale smooths out shorter-term variations in star formation \citep{kim26}, which cannot be captured by our technique (even though we can measure sSFR up to 60~Gyr$^{-1}$; Fig.~\ref{fig:simu}). In practice, SED fitting integrates emission over timescales that are too long to resolve such rapid events. These variations emerge when adopting a shorter $t_0$, such as 10 Myr (setup {\it b)}, Appendix~\ref{app:fitsystematics}), at least up to $z\sim9$. However, even this setup fails to recover extreme sSFR at higher redshift. 
We cannot exclude 
that early SFHs may be characterised by even shorter bursts, beyond the reach of our method.

We examined different systematics potentially responsible for the non-evolving sSFR. 
First, we verified that the overall trend is robust against  SED fitting assumptions (Appendix~\ref{app:fitsystematics} and Fig.~\ref{fig:ssfr_median}), despite individual galaxies may be strongly affected,  as can the median SFR and stellar mass. 
Secondly, we considered observational incompleteness. 
The $10^8-10^9 M_\odot$  mass range is  
characterised by a $>$90\% level completeness at least out to $z\sim 11$ (Sect.~\ref{sec:completeness} and Fig.~\ref{fig:simu}).  
Additionally, the SC-SAM, which is mass-matched to our sample and thus affected by a similar level of mass incompleteness, shows an increasing trend (dark red dashed line in Fig.~\ref{fig:ssfr}) that is not seen in our data. 
We concluded that mass incompleteness 
cannot explain the flattening, 
unless a critical population of galaxies with high sSFR is completely missing from both the observed sample and the simulation used to quantify the incompleteness (see Sect.~\ref{sec:missing}). 
Finally, we verified that the  lack of optical rest-frame constraints at high redshift can account on average for 0.1~dex (Fig.~\ref{fig:check}), 
which is well within the scatter of the sample and not sufficient to reproduce the increasing trend predicted by most theoretical models and discussed in Sect.~\ref{sec:models}. 

\begin{figure*}[t!]
    \centering
\includegraphics[width=1.0\textwidth]
{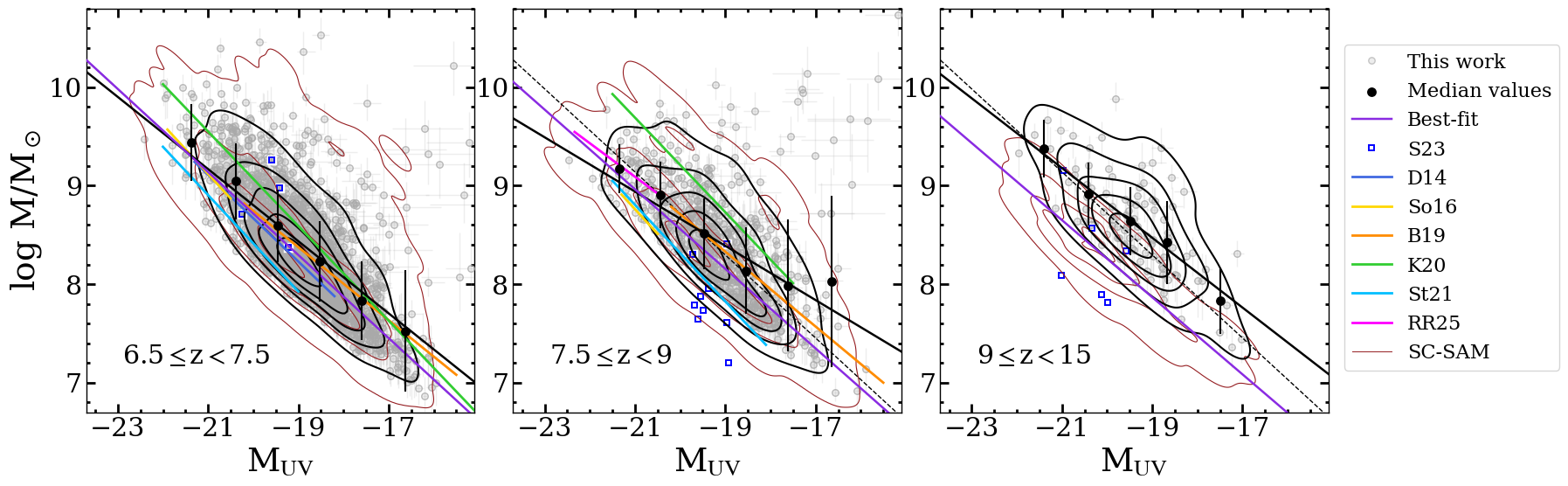}
\caption{Relation between the stellar mass and the rest-frame UV absolute magnitude 
in three redshift bins. Thin black lines enclose 10\%, 30\%, 50\%, 70\%, and 90\% probability densities. Large black symbols represent the median mass and 16${th}$--84${th}$ percentile range in bins of absolute magnitude.  
 The black  line shows the best-fit relation, 
 and the  dashed lines replicate the best fit at $z\sim 7$ in the other redshift bins.  The coloured lines and symbols illustrate previous results from the literature: \citet[][D14]{duncan14}, \citet[][So16]{song16}, \citet[][B19]{bhatawdekar19}, \citet[][K20]{kikuchihara20}, \citet[][St21]{stefanon21}, \citet[][S23]{santini23}, \citet[][RR25]{rojas-ruiz25}.  The dark red thin curves enclose 10\%, 50\%, 80\%, and 99.9\% 
probability densities for the SC-SAM. 
}
\label{fig:masslum}
\end{figure*}

\subsection{A population of galaxies with low SF activity} \label{sec:passive}

Remarkably, we observe a population of galaxies hosting very low star formation activity up to very early cosmic epochs ($z\sim10$). 
They are characterised by low SFR ($\lesssim  3 M_\odot/yr$, but typically $<0.5$), masses at the high end of the distribution (mostly \lgm~$>9$), and show a wide range of dust attenuation. 

The presence of quiescent galaxies at the redshifts probed by our analysis has already been reported by previous works by means of dedicated photometric searches \citep[e.g.][]{merlin25,russell25,baker25} as well as spectroscopic confirmations \citep[$z\sim 7.3$,][]{looser24,weibel25}.  
We compared our results with our previous analysis \citep{merlin25}, where we identified 12 candidates at $z>6.5$ with \lgm~$>9.5$ adopting a different SED fitting and selection technique. To this aim, we consider the most extreme examples in our present sample, i.e. the 14 galaxies with sSFR lower than 0.1 Gyr$^{-1}$.  Six of them have negligible (4 sources) or very low sSFR ($\sim 0.15-0.25$ Gyr$^{-1}$, 2 sources) in \cite{merlin25}, including an object at $z\sim 10.2$, although none of them  eventually passed the final, conservative selection in our previous work (sometimes only because below the chosen mass limit). However, among our 14 quiescent candidates 
are object that are fitted as star-forming galaxies by \cite{merlin25}. This comparison 
highlights how sensitive the results are to the selection criteria used to identify this class of galaxies. 
Finally, we note that the two spectroscopically confirmed quiescent galaxies by \cite{weibel25} and \cite{looser24} are fitted as quiescent in the present work, but just above the chosen threshold  (sSFR$\sim$0.1 and 0.4 Gyr$^{-1}$).

While a detailed  analysis of these candidates and investigation of differences among various selection techniques are beyond the scope of the present paper, 
our findings are in broad agreement with a scenario where a fraction of massive galaxies quench at early times. This illustrates the variety of SF conditions already in place as early as $z\sim 10$.

\subsection{The mass-luminosity relation} \label{sec:masslum}

We explore here the stellar content of our galaxies 
by looking at the relation between the stellar mass and the rest-frame UV luminosity in three redshift bins, shown in Fig.~\ref{fig:masslum}. 
Thanks to the large dataset, spanning a wide sky area and probing low luminosities, our sample extends beyond previous analyses, covering five magnitudes in luminosity and over three dex in stellar mass. 

We do observe a moderate to high correlation  between the stellar mass and the absolute magnitude  
(Pearson coefficient between -0.5 and -0.7,
p-value $\ll 0.001$), and fit 
a mass-luminosity relation which 
is consistent with no evolution over the redshift range probed by the present work, 
as shown by the best-fit parameters reported in Table~\ref{tab:masslum}.  
While the relation between mass and UV luminosity has been shown for spectroscopic $z=9-14$ galaxies by \cite{tang26}, we note that to our knowledge this is the first attempt to derive 
a best fit 
at these redshifts. 

The scatter around the relation is 
very large: for a given luminosity, the bulk of the population  spans one order of magnitude in stellar mass, but individual galaxies 
can vary by more than 
a factor of 100, reflecting a large variation in the stellar population properties. For this reason, we warn against the use of the UV luminosity as a tracer of the stellar mass, a common practise until the advent of JWST due to the lack of IR constraints (see extended discussion in \citealt{santini23}).

We compared our results with previous works. We first considered our previous analysis \citep{santini23}, where we observed no correlation between the stellar mass and the luminosity. This was likely a consequence 
of the small number statistics affecting the very first JWST data, only probing a limited range in luminosity. Additionally, our previous results were likely prone to the outshining effect, known to lead to underestimated stellar masses \citep{papovich01,leja22}. We now infer larger masses at $z>7.5$ due to the adoption of more flexible SFHs, which include the possibility of a recent burst to reproduce the UV light (either through bins of constant star formation in the non-parametric fits, or through an exponential burst or a constant value prior to the observation for parametric SFHs). Moreover, thanks to the possibility of explaining the UV light with a recent burst of star formation, we could fix the minimum age of the oldest stellar populations to 100 Myr instead of 10 Myr, which in  our previous analysis was responsible for the very low mass measurements. 
At $z\sim 8$, our best-fit relation nicely agrees with the estimate of \cite{rojas-ruiz25}, derived from bright BoRG-JWST \citep{roberts-borsani25} galaxies. Our results also overlap with those from \cite{bhatawdekar19}, based on deep HST photometry in the Hubble Frontier Fields and extending to low masses and luminosities. However, compared to other pre-JWST works, the best-fit relation is offset in both directions (with \citealt{stefanon21}, \citealt{duncan14} and \citealt{song16} finding a lower normalisation and \citealt{kikuchihara20} a larger one), with the offset being more pronounced at $z\sim 8$ than at $z\sim 7$. This highlights the importance of deep, rest-frame near-IR photometry for estimating accurate stellar masses at high redshift.

We also compared our results with the mass-matched SC-SAM.  At $z<9$ the data are well reproduced, with the exception of the most extreme sources with large stellar masses and faint luminosities. In both redshift bins, the  mass–luminosity relation inferred from the SC-SAM agrees with the one observed at $z\sim7$. At $z>9$, instead, it is lower by a factor of $\sim$3, with the SC-SAM predicting brighter luminosities for a given stellar mass. We note, however, that the simulation nearly lacks sources with $M\gtrsim 10^9M_\odot$ at these redshifts. One possible explanation for this mismatch in luminosity is the  absence of dust reddening in the SC-SAM at $z\gtrsim10$.  
In summary, the comparison 
suggests that the model is unable to reproduce the full variety of galaxies, as discussed in the next section.

\subsection{The diversity of the stellar populations} \label{sec:stellarpop}

\begin{figure}[t!]
    \centering
\includegraphics[width=13.5cm]{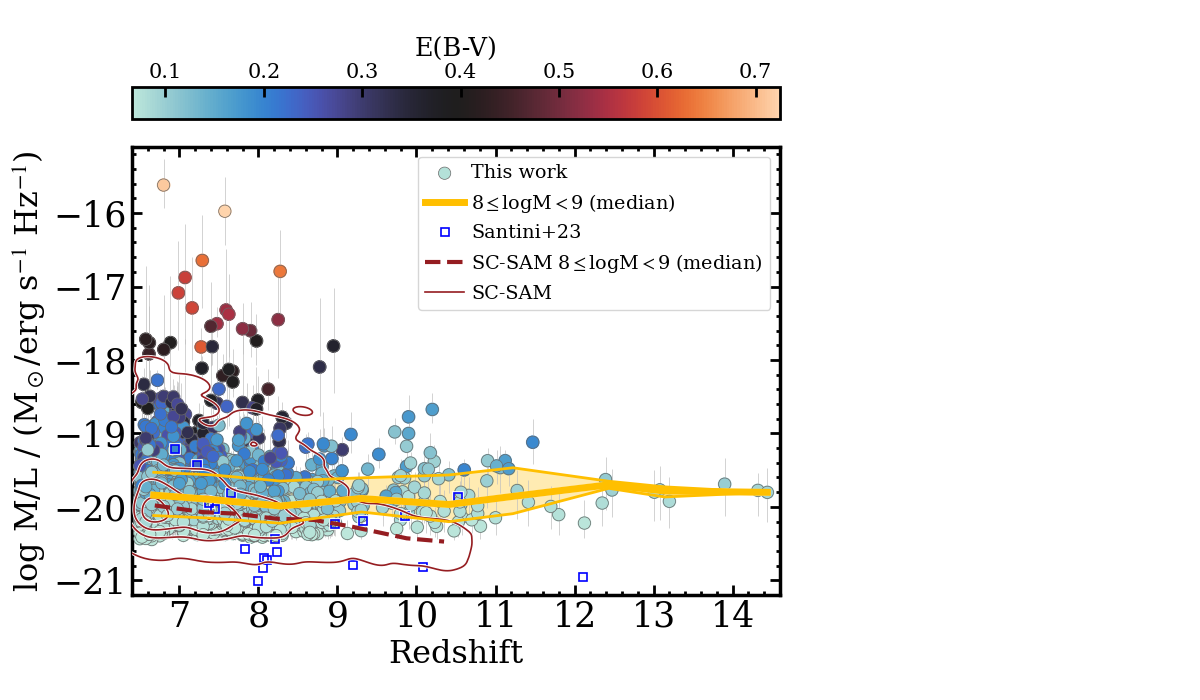}
\caption{Evolution of the \ml, where L is the rest-frame UV luminosity. Points are colour-coded according to the dust attenuation \ebv.  
The thick yellow line indicates the median   of the \lgm~$=8-9$ subsample and the shaded region spans the 16${th}$--84${th}$ percentile range. 
The dark red dashed line and solid thin curves are the median and 10\%, 50\%, 80\%, and 99.9\% 
probability densities for the SC-SAM.  
}
\label{fig:ml}
\end{figure}

We finally focused on the mass-to-light ratio, which is a proxy of the stellar populations hosted by the galaxy. 
In Fig.~\ref{fig:ml} we report the redshift evolution of \ml. 
As discussed in Sect.~\ref{sec:masslum}, we derive larger \ml~compared to our previous work thanks to mitigation of the outshining effect on the inferred stellar masses.

For a given redshift, our sample spans up to four orders of magnitude in \ml~at $z\sim7$ and roughly a factor of 50 at $z\sim10$, irrespective of the fitting setup (see Fig.~\ref{fig:ssfr_median}). 
We note that spectroscopic sources span the  whole distribution, indicating that potential photometric redshift degeneracies do not impact our conclusions.

Notably, we observe high  \ml~ratios  up to $z\sim10$. 
In principle, a high \ml~indicates the presence of evolved stellar populations, although dust attenuation can increase the observed value even in young, star-forming galaxies. 
High-\ml~sources have large stellar masses (\lgm~$\gtrsim9$) and are typically heavily attenuated (\ebv~$>0.3$).
They however have 
ages larger than 350 Myr, 
occupying the high-age tail of the distribution. 
Unsurprisingly, a fraction (26\%) of the most extreme ones (23 sources with log(\ml) $>-18$, where \ml~is expressed in units of $M_\odot$/(erg s$^{-1}$ Hz$^{-1}$))  
overlap with the population of quiescent galaxies discussed in Sect.~\ref{sec:passive}. The remainder have sSFR typically below $\sim$1/Gyr, with only two sources exceeding 2/Gyr. Although some may still be forming stars at a moderate rate, these galaxies have already assembled a large stellar mass and likely 
host 
evolved stellar populations. Conversely, roughly half of the most extreme quiescent systems have log(\ml)~between $-18$ and $-19$ (in the same units) and low dust attenuation (\ebv~$<0.3$), suggesting that they recently stopped forming stars and still retain light from comparatively younger stars. 

These results indicate that the high-redshift galaxy population is largely diverse, as also suggested by the scatter around the mass-luminosity relation. 
Galaxies are observed at different evolutionary stages, and their diversity reflects a wide range of star formation histories and dust attenuation, 
as indicated by the colour-coding of Fig.~\ref{fig:ssfr} and \ref{fig:ml}. 
In particular, we can identify distinct galaxy classes: {\it a)} star-forming sources with moderate level of dust attenuation (\ebv~$<0.2$) and young stellar populations, covering the full redshift range and accounting for $\sim75\%$ of the sample; {\it b)} more dusty (\ebv~$\sim0.2-0.7$) star-forming galaxies at $z\lesssim 9-10$;   {\it c)} galaxies with low sSFR spanning the whole \ebv~range ($\sim 0.1-0.7$)  and characterised by intermediate to large \ml. 
The observed variety in early galaxy properties confirms with much larger statistics the results obtained by our previous analysis \citep{santini23} as well as by other studies in the literature \citep[e.g.][]{dressler24,conselice25,tang26}.

As discussed in Sect.~\ref{sec:masslum}, the diversity observed in galaxy stellar populations is not fully reproduced by theoretical simulations, which struggle to produce high \ml~and evolved stellar populations 
at high redshift  \citep[see][and references therein]{merlin25}, 
especially at $z>9$. 
This likely reflects the smooth, monotonic SFHs in SAMs \citep{cote18}: the lack of intense bursts or rapid growth phases reduces the scatter in M/L and hampers the formation of relatively evolved stellar populations at early times.

Even though galaxies with high \ml~ratios become rarer and rarer at earlier times, as expected, the median observed \ml~calculated on $10^8-10^9$ M$_\odot$ galaxies does not show any redshift evolution. While mass incompleteness may  play a role, this cannot entirely explain the observed trend for two reasons. 
First, we have shown that the level of completeness 
 is very high in that mass range (see Sect.~\ref{sec:completeness}). 
 Secondly, as discussed in Sect.~\ref{sec:ssfr}, 
mock galaxies extracted from the SC-SAM show a decreasing trend despite being 
mass-matched to our sample and therefore  
suffering from a similar level of mass incompleteness. 
Finally, the observed flat \ml~cannot be explained by SED fitting systematics, 
as different fitting setups provide the same picture (see Appendix~\ref{app:fitsystematics} and Fig.~\ref{fig:ssfr_median}) nor by the lack of optical constraints at $z>10$ (Appendix~\ref{app:accuracy} and Fig.~\ref{fig:check}).

\section{Discussion} \label{sec:disc}

We attempt here to interpret our findings by  comparing them with theoretical predictions and exploring how galaxies have assembled their mass.  
Before proceeding, 
we stress that our conclusions on the galaxy  assembly histories, progenitors and their observability are speculative and entirely stand on the ability to correctly recover galaxy SFHs, 
known to be an extremely challenging task \citep[e.g.][]{haskell24,wang25,carvajal-bohorquez25}.  For this reason, we repeated the analysis  by considering different SFH assumptions (namely setup e) and setup f) described in Appendix~\ref{app:fitsystematics}), and found  very similar results.   
We also note that we considered the best-fit SFHs, whose associated properties may slightly vary compared to the median values adopted in this work, and assume that each galaxy derive from one single progenitor.

\subsection{The theoretical framework} \label{sec:models}

Since the first JWST results appeared, theoretical models have struggled to reproduce the many unexpected findings, the most established one being the excess of UV-bright galaxies at 
$z\gtrsim10$. Different models have implemented  different prescriptions to address this issue. As detailed in the introduction, three main classes of interpretations can be identified: models boosting the UV luminosity, models enhancing the star formation efficiency, 
and models invoking a change in the cosmological paradigm responsible for accelerating galaxy growth, as suggested by recent DESI results \citep{adame25}. We compare here the observed evolution of the sSFR with 
examples from each class above.

We first considered the Attenuation Free Model (AFM hereafter) of \cite{ferrara23}. It explains the excess of UV luminosity by reducing dust attenuation through radiation-driven outflows that efficiently expel dust from early galaxies, enhancing visibility. These outflows are developed once the galaxy luminosity becomes super-Eddington, i.e. once galaxies exceed a given threshold in sSFR \citep{ferrara24a}.  

We then considered two variants \citep{cantarella26} of the GAEA semi-analytical model \citep{delucia24} aimed at enhancing star formation by suppressing the feedback. The first one is the 
`feedback-free starburst' (FFB) model, inspired by \cite{dekel23}, combined with a weakened feedback.  This variant increases the star formation efficiency at $z > 10$  within the densest cold gas regions ($\gtrsim 10^3 cm^{-3}$) by shortening the free-fall time to 1 Myr,  
below the time required for stars to develop winds and supernovae. 
In addition, the stellar feedback is reduced at $z>4$ (see \citealt{cantarella26} for detail). 
The second model is the ‘no high-z stellar feedback’ variant, in which the stellar feedback is suppressed  across all galaxies  
at $z>10$ by shutting down gas reheating and energy injection. 

Finally, we examined the framework developed by \cite{menci24} to test the effect of assuming a Dark Energy with negative cosmological constant $\Lambda$. A negative $\Lambda$ boosts the dark matter halo masses at high redshift with respect to the $\Lambda$CDM scenario ($\Lambda=0.7$), therefore naturally produces higher stellar masses and higher luminosities, which are obtained by standard assumptions and maximising the SF activity. To highlight this effect, we considered the extreme case of $\Lambda=-2$. 

\begin{figure}[t!]
    \centering
\includegraphics[width=13cm]{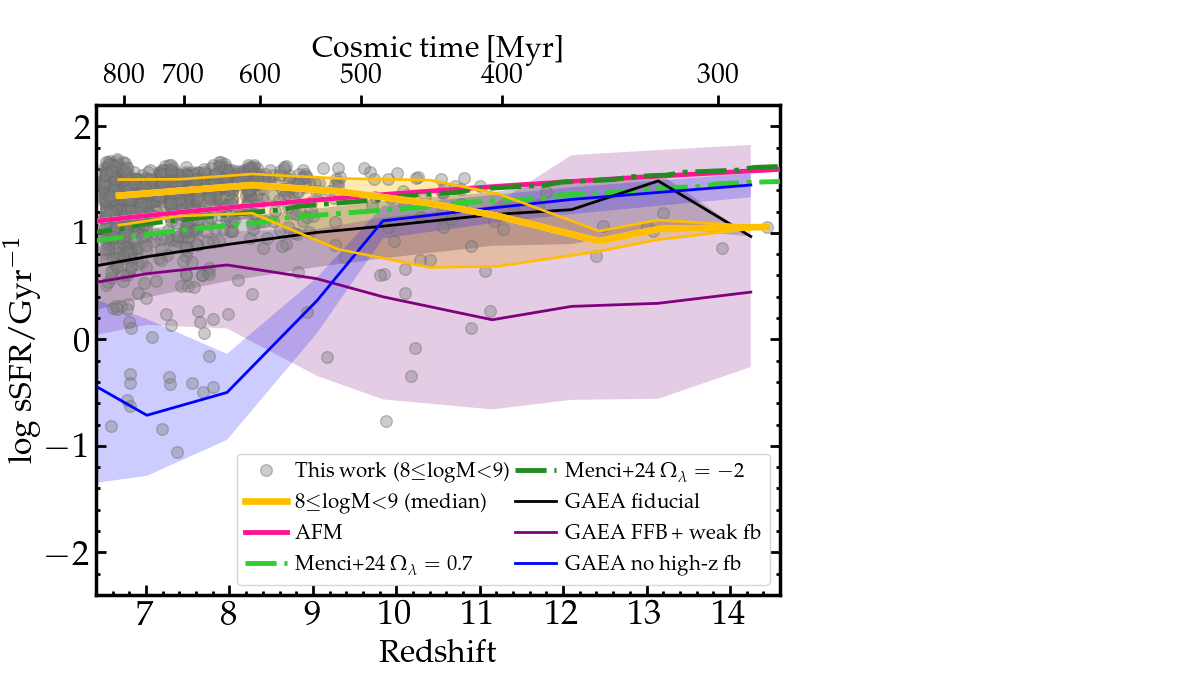}
\caption{Evolution of the sSFR. The grey symbols are the observed galaxies in the $10^8-10^9 M_\odot$ mass range, and the yellow line and shaded region indicate the median and 16${th}$--84${th}$percentile range. 
The magenta line shows the prediction of the AFM model of \cite{ferrara24a}. The dash-dot light and dark green curves are the expectation of the framework outlined by \cite{menci24} for $\Lambda=0.7$ and  $\Lambda=-2$, respectively. 
We show in black, purple, and blue the expectation of the  GAEA fiducial model, `FFB + weak feedback', and `no high-$z$ feedback' variants, respectively; the solid lines and shaded regions represent the median and the 116${th}$--84${th}$ percentile range of $10^8-10^9 M_\odot$ model galaxies. 
}
\label{fig:models}
\end{figure}

In Fig.~\ref{fig:models} we show the predicted redshift evolution of the sSFR according to the models described above, compared to the observed evolution of $8\leq$ \lgm~$<9$ galaxies.  

According to the AFM, the sSFR increases with redshift as $(1+z)^{3/2}$ (magenta line in Fig. \ref{fig:models}, see also the similar trend $\propto (1+z)^{2.25}$  predicted by  \citealt{dekel13,dave11}). A similar behaviour is shown by the fiducial GAEA model (black line and shaded region) and by the framework developed by \cite{menci24} (dark and light green dash-dot curves). In the latter case, a negative $\Lambda$ only results in a slightly larger normalisation, implying that it boosts both SFRs and stellar masses in a rather similar way  at all redshifts. All these trends are in contrast with the 
lack of evolution 
shown by the observed galaxies. Interestingly, the two GAEA model variants exhibit opposite behaviours, as also highlighted by the authors \citep[see][and their Fig.~11]{cantarella26}, due to the combined effect of free-fall time and feedback, responsible for gas consumption and depleting, respectively. 
In the `FBB + weak feedback' model, the gas reservoir is rapidly converted into stars at early times  
in a bursty mode, leading to a milder star formation activity at the redshifts probed in this work. The latter is sustained over time by the stellar 
feedback, which prevents extremely high sSFR and at the same time a too early gas consumption.  
Although  roughly 0.7 dex lower in normalisation and unable to reproduce galaxies with high sSFR at $z\lesssim 9$, only this model variant depicts a redshift evolution which is similar to the observed one. We note that it also reproduces the quiescent galaxies when the full 3$\sigma$ distribution is considered \citep[see also][]{dekel25}. 
In contrast, in the `no high-$z$ stellar feedback' scenario the gas amount is larger at early epochs, leading to a larger SFR, which is suppressed at later times ($z<10$) when feedback becomes effective. While this scenario is capable of reproducing the observed $z\gtrsim10$ galaxies and sources with low sSFR at $z\sim 7-8$, it does not reproduce the bulk of the population at $z<9$. A similar result is found by
 \cite{cantarella26}, who demonstrate how this model variant is in contrast with recent works presenting the Main Sequence at $z\sim 7-8$ \citep{leethochawalit23,cole25}. 

The overall theoretical picture remains unclear. Despite several mechanisms have been proposed to match the observed $z>10$ luminosity functions, yet none appears able to reproduce the full diversity of galaxies, suggesting that there is no single explanation and that a combination of physical processes is likely at play.
In this context, the sSFR seems to be a powerful  quantity for  discriminating among the various interpretations, with different models predicting different evolutionary trends. However, in order to use the observed sSFR evolution as a benchmark, we need to exclude any possible observational bias affecting our results.

\subsection{Are we missing a substantial population of starbursting galaxies?} \label{sec:missing}

As shown by the comparison with model predictions (Sect.~\ref{sec:models}), assessing the evolution of the sSFR is crucial 
to understand the physical mechanisms regulating the formation and evolution of the first galaxies. 
The observed median sSFR and \ml~show no 
evolution over $\sim 600$ Myr of cosmic time.  Is this the true evolution - at odds with 
most models except  
the `FFB + weak feedback' scenario implemented in the GAEA simulation \citep{cantarella26} - or 
does our dataset miss a population of highly star-forming galaxies at high redshift? 
If such galaxies exist, they would have high sSFRs and young stellar populations, and thus low \ml. Their absence from our sample would push the median sSFR (\ml) towards lower (higher) values, potentially explaining the 
mismatch between the observed redshift evolution and that predicted by most models.

\begin{figure*}[t!]
    \centering
\includegraphics[width=9cm]{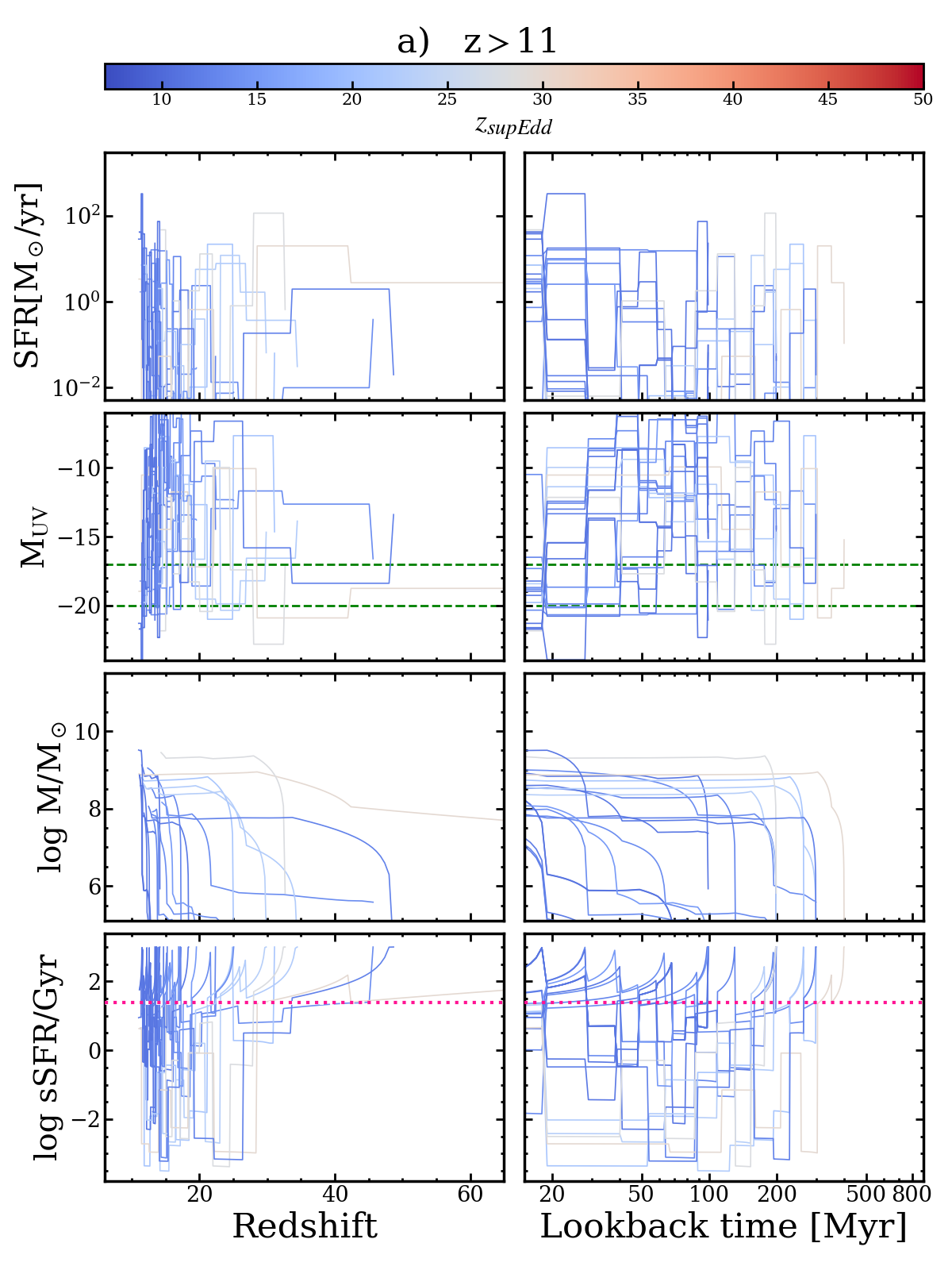}
\includegraphics[width=9cm]{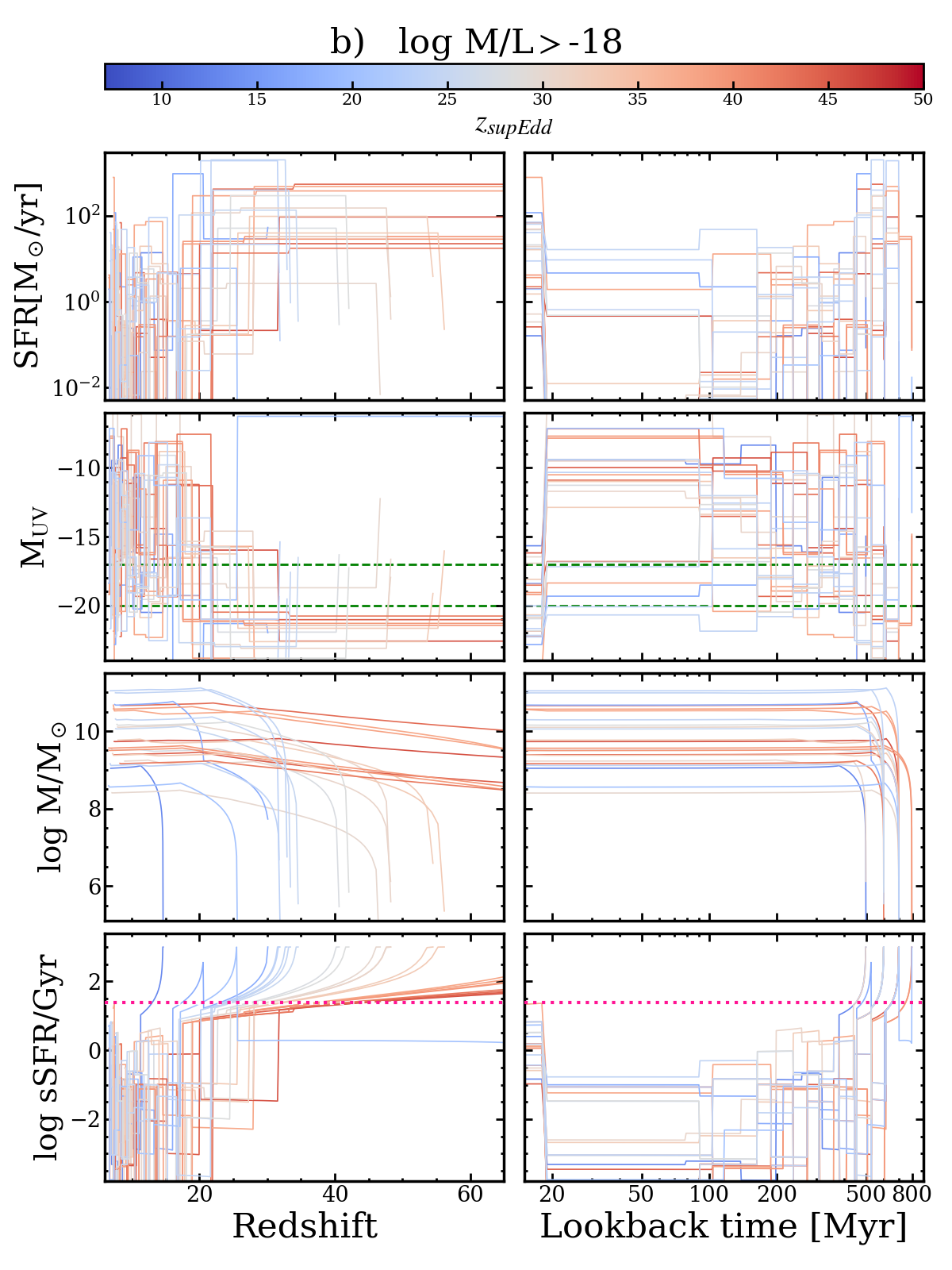}
\caption{{\it Panel a)} SFR ({\it top}), UV absolute magnitude ({\it second row}), stellar mass 
({\it third row}), and sSFR 
({\it bottom}) as a function of redshift ({\it left}) and lookback time ({\it right})  of $z>11$ galaxies. 
{\it Panel b)} Same as panel {\it a)} but for 
$\rm \log (M/L)$$/(M_\odot/(erg~s^{-1}Hz^{-1})>-18$
galaxies, mostly concentrated between $z\sim 7$ and $z\sim 8$.   
The curves are colour-coded according to the redshift when  the corresponding galaxies last crossed the outflow threshold, shown by the horizontal magenta dotted line; 
in the context of the AFM, the sources shown in blue have recently experienced a phase of elevated sSFR, while those in red  are in the sub-Eddington phase since $z\sim 30$ at least. 
The green lines in the second row panels show the depth limits in UV absolute magnitude of -17 and -20. 
}
\label{fig:tracks}
\end{figure*}

In Sect.~\ref{sec:completeness} we  tested our sensitivity to a potential population of starbursting, extremely dust-enshrouded sources at early epochs. As discussed above and shown in the right panels of Fig.~\ref{fig:simu}, our selection technique is unable to include dusty galaxies at $z>10$, so we cannot exclude their existence.  
As a matter of fact, none of the 
$z>9$ (11.5) galaxies in our sample are fitted with \ebv~larger than 0.2 (0.1), 
at variance with $z<9$ sources, where a small fraction of objects shows moderate or high levels of dust obscuration 
(up to \ebv~$\sim0.7$, see Fig.~\ref{fig:ssfr} and \ref{fig:ml}). 
The lack of optical constraints at high redshift makes the selection more prone to dust obscuration issues. 
While extremely dusty (\ebv$\gtrsim$0.5, or $A_V$$\gtrsim$2) galaxies at these redshifts have not been spectroscopically confirmed yet, dust attenuations as high as $A_V\sim 0.8-1.2$ at $z\sim10-12$ have recently been  measured  \citep{mitsuhashi25,donnan25,rodighiero26} 
suggesting that dusty galaxies may indeed exist at such early epochs. 

The existence of a dusty, starbursting galaxy population has been postulated within the framework of the AFM %
\citep{ferrara24a}.  
According to this model,  
at $z\sim10$ the typical sSFR exceeds a critical value (shown by the dotted magenta horizontal line on Fig.~\ref{fig:ssfr}) above which galaxies develop the 
outflow capable of heating and expelling dust and gas. 
This super-Eddington outflow phase,  
dubbed blue monster phase, is preceded by an 
intense, highly dust-obscured phase of star formation, during which 70\% of stars are formed, claimed to be unobservable with JWST due to extreme attenuation \citep{ferrara24b}. 
In this context, given the outcome of our simulation (Sect.~\ref{sec:completeness} and Appendix~\ref{app:completeness}), extremely dusty galaxies would be missed, and the only star-forming phase accessible to our study would therefore be the dust-free one.  The fact that 
none of the galaxies at $z\gtrsim 12$ is in or close to the super-Eddington regime (i.e. above the given threshold in sSFR), may imply that the duty cycle of the blue monster phase at these early epochs must be shorter than $10$–$20$ Myr,\footnote{A similar result is obtained when adopting $t_0=10$ Myr; see setup {\it b)} in Appendix~\ref{app:fitsystematics}.} hence shorter than the $\sim$40 Myr predicted by \cite{ferrara24b} for the case of GS-z14-0 \citep{carniani24}. 
The short duty cycle, combined with limited statistics at extreme redshifts, may determine our failure in capturing 
galaxies in the dust-free phase following the outflow and preceding the SF suppression due to gas removal. 

\subsection{The assembly history of the highest-redshift galaxies } \label{sec:assembly}

Although we cannot directly probe the existence of $z>10$ dust-obscured galaxies with current data, 
we can test the hypothesis that the highest-redshift sources in our sample 
result from the suppression of star formation in galaxies that have recently undergone an intense star-forming phase.  
To this aim, we traced back the SFH reconstructed from SED fitting to assess whether these galaxies experienced a higher SF in their recent past. 
We calculated their mass assembly history, accounting for the recycle fraction, and thus derived their sSFR histories (sSFHs).  Peaks in the sSFH correspond to phases of major mass assembly.

For the sake of clarity, we limited the analysis to the most distant sources at $z>11$ (19 sources - a similar scenario is obtained when considering all $z>10$ galaxies), and show the results 
in the left panel of Fig.~\ref{fig:tracks}. 
Their SFHs and sSFHs are variegated and show ups and downs, similar to the so-called lulling or breathing galaxies \citep[e.g.][]{looser25,gelli25,merlin25}. 
Fourteen  of these galaxies (74\%) have experienced a phase of intense growth - and  
had a sSFR above the magenta dotted line - 
over their past $\lesssim$50 Myr, 
populating the super-Eddington regime for short, repeated episodes during that time.  
This fraction is 
consistent with the fraction of super-Eddington galaxies predicted by the AFM at these redshifts, i.e. $\sim$0.5 at $z\sim 11$ and $\sim$0.75 at $z\sim 14$ \citep{ferrara24a}, qualitatively in agreement with the scenario depicted by the model.   
The remaining five 
galaxies have 
SFH peaking at $z\sim 20-40$, and have not been experiencing super-Eddington sSFR since $z\sim 20-30$. 
The entire sample of $z>11$ galaxies were formed and have assembled their entire mass 
over their past 150-300 Myr, 
indicating that we are witnessing the process of formation of these systems.

By making reasonable assumptions, it is possible to assess the observability of the progenitors of the highest-redshift galaxies. 
Starting from the SFH, 
we estimated the UV luminosity as a function of cosmic time by converting the SFR through the typical SFR-to-UV luminosity ratio ($K_{UV}^{\rm obs}$) observed for our dataset. As expected, $K_{UV}^{\rm obs}$ is correlated with dust attenuation, with larger values observed for higher levels of obscuration. The values of $\log K_{UV}^{\rm obs}$  span 3 dex, but 85\% of the sample lies between $-27.8$ and $-27.1$ $M_\odot$yr$^{-1}$/(erg s$^{-1}$Hz$^{-1}$), with the distribution peaking at $-27.7$. At the highest redshift ($z\gtrsim12$), where we are interested in reconstructing the UV luminosity, 
the distribution of
$\log K_{UV}^{\rm obs}$ is much narrower and limited to the range between  $-27.7$ and $-27.85$. As a first approximation, we adopted  $\log K_{UV}^{\rm obs}=-27.7$, keeping in mind that a 0.4 dex change in $\log K_{UV}^{\rm obs}$ reflects into a change in the UV luminosity of 1 magnitude. 

The resulting luminosities at different times are shown in the second row panels of Fig.~\ref{fig:tracks} (left panels). 
The progenitors of the highest-redshift sources typically have faint UV luminosities. Only $\sim 20\%$ of galaxies are brighter than a magnitude of $-20$ (the threshold for bright objects, i.e. more or less the  knee of the luminosity function, \citealt{finkelstein24}) beyond $z \sim 15-20$, 
and only
$\sim 35\%$ exceed $-17$, which is the expected 
faint-end limit the search can reach 
\citep{perez-gonzalez25}.
Overall, the majority of the progenitors of the highest-redshift galaxies are expected to remain hidden to our view even without the need of an extremely dust obscured phase. We note  that the fraction of observable sources would further decrease in case of high dust attenuation.

In summary, within the framework outlined by the AFM,  
our data are consistent with a scenario in which we are observing the highest-redshift galaxies shortly after the onset of radiation-driven outflows and at the conclusion of a dust-enshrouded star formation episode, in a phase of more moderate activity. The most intense phases of formation of these galaxies cannot be observed with the available data. 
Similar conclusions on the past history of high-redshift galaxies have been proposed by \cite{dressler24} to explain the lack of sources with increasing SFH as recovered by SED fitting, and by \cite{nakazato25} based on theoretical arguments.

\subsection{The progenitors and assembly of evolved galaxies 
}\label{sec:progenitors}

Here we focus our attention on another peculiar subclass of galaxies, i.e. those galaxies that show high \ml~within the first few hundreds megayears after the Big Bang. 
Again, we limit the analysis to the most extreme sources,  showing those with the highest \ml, i.e. 
log(\ml) $>-18$, where \ml~is expressed in units of $M_\odot$/(erg s$^{-1}$ Hz$^{-1}$),
mostly observed at $z\sim 7-8$.  
We explore their assembly histories and make predictions on the observability of their progenitors at different cosmic times. 
We show the results in the right panels of Fig.~\ref{fig:tracks}. 

The vast majority of these galaxies formed the bulk of their stars 500–800 Myr before observation, i.e. at $z\gtrsim20$–30. By $z\sim11-14$ they had already assembled (most of) their final stellar mass, exceeding the masses of the galaxies observed at those redshifts, which therefore are unlikely to be their progenitors. 
Galaxies with high \ml~ experienced  
intense SF activity in their past, and are characterised by SFH that are much more self-similar than those of the highest-redshift sources, which are still in the process of assembling and show a high level of stochasticity. 
Despite being dusty (Sect.~\ref{sec:stellarpop}), their assembly histories suggest that their high \ml~indicate evolved stellar populations, which are already in place at these early cosmic times \citep[see Sect.~\ref{sec:passive};][]{looser24,kuruvanthodi24,russell25,merlin25,baker25}. This result provides further 
evidence for the hypothesis that 
baryonic mass assembly might be occurring more rapidly than current galaxy formation and cosmological models predict \citep[e.g.][]{liu22,menci22,dekel23,melia23,padmanabhan23}.

The progenitors of evolved galaxies at $z\sim 7$-8 are expected to be luminous ($M_{UV}<-20$) at $z>20$, i.e. during the epoch of their formation. Despite the rarity of these sources, which represent $\sim1\%$ of our sample, finding bright candidates at extremely high redshift \citep[as could be the case of Capotauro,][]{gandolfi26}
should not be unexpected. We note, however, that we may expect the phase of major assembly of these galaxies to be dust enshrouded, causing a different SFR-to-UV luminosity conversion. Considering a conservative case, i.e. $\log K_{UV}^{\rm obs}\sim -26$, corresponding to an \ebv~of $\sim 0.6$, only 22\% of the sources would exceed the $-20$ limit, but 74\% of them would still be brighter than $-17$, therefore within reach of our telescopes.

For reference,  in Fig.~\ref{fig:tracks_lowssfr} we show the assembly histories of  quiescent galaxies   with sSFR $<$ 0.1 Gyr$^{-1}$. 
Some of them, typically the most massive ones, have formed their stars very early, and they did not experience significant SF over the past 500 Myr at least (red to light blue coloured curves). Conversely, 
the least massive ones recently underwent a phase of super-Eddington SF over the past 300 Myr (bluish curves), consistently with a downsizing scenario \citep{cowie96,fontanot09} and 
qualitatively in agreement with the results of \cite{merlin25}, therefore still host relatively young stellar populations. Examples of young stellar populations in  galaxies with low current SF, i.e. experiencing bursty star formation, are also reported by \cite{looser25}.

\section{Summary and conclusions}\label{sec:summary}

We investigated the specific SFR (sSFR) and the stellar populations as traced by the mass-to-light ratio (\ml) in a sample of 2420 $z=6.5-14.5$ galaxies carefully selected from the ASTRODEEP-JWST database adopting photometric and photo-$z$ criteria. We modelled this sample assuming non-parametric SFHs and extensively tested the robustness of our results against systematics related with the SED fitting assumptions, in particular the choice of the SFH. After validating the inferred physical properties against spectroscopic measurements and assessing the robustness of stellar masses and sSFR in the absence of optical constraints at $z\gtrsim10$, we obtained the following results: 
\begin{itemize}
    \item[$\bullet$] The sSFR of the bulk of the population shows no evidence of evolution with cosmic time at $z>6$, as indicated by the flat median trend across the redshift range probed.
    \item[$\bullet$] We measured a non-evolving 
    relation between the stellar mass and the rest-frame UV luminosity in three redshift bins, 
    with a large scatter of more than two orders of magnitude in mass at a given luminosity.
    \item[$\bullet$] The mass-to-light ratio spans up to four orders of magnitude at $z\sim 7$ and a factor of $\sim$50 at $z>10$, indicating a wide diversity in galaxy stellar populations, and its median shows no evolution with cosmic time.
    \item[$\bullet$] We identified distinct galaxy populations: 
    star-forming systems with moderate (\ebv~$<0.2$) dust attenuation observed across the full redshift range; 
    more heavily obscured (\ebv~$\sim 0.2-0.7$) star-forming galaxies predominantly at $z \lesssim 9-10$; 
    quiescent galaxies (sSFR $<$ 0.1 Gyr$^{-1}$) in place since $z \sim 10$. Remarkably,  
    galaxies with high \ml, observed up to $z \sim 10$, include a mixture of the latter two populations. 
\end{itemize}

None of the theoretical predictions analysed in this work can reproduce the observed diversity of physical properties, suggesting that a combination of mechanisms is required to explain the data. The evolutionary trends in our measurements shows  qualitative  agreement with the scenario of the FFB model combined with weak feedback at high-$z$ implemented in the GAEA simulation \citep{cantarella26},  
and are in contrast with the  increasing  
trend of the sSFR   
towards earlier epochs expected by the other models.  
However, this mismatch can be explained by the presence of a population of intense star-forming, highly dust obscured galaxies hidden to our view. The existence of such a population has been  postulated by the 
Attenuation Free Model (AFM) of \cite{ferrara24a} as a phase of major mass assembly preceding 
a radiation-driven outflow that clears dust and gas. 
While the dusty phase would be out of reach  with 
current data, 
the so-called blue monster phase, immediately following the outflow and prior to the 
SF suppression, would  not be captured by our observations likely because of its short duty cycle  at very high redshifts ($z>12$). 

Although we cannot observe $z>10$ dusty galaxies with our data, we can attempt to corroborate their existence by indirect means. 
Tracing back the SFH and making assumptions on the SFR-to-UV luminosity ratio, we attempted to recover the assembly histories and  
observability of the progenitors of a few  
extreme classes of objects.  Under the assumption that the SFHs are correctly recovered, we  speculate the following:  
\begin{itemize}
\item[$\bullet$] The majority of the highest-redshift ($z>11$) galaxies have experienced a vigorous phase of SF shortly ($\lesssim 50$ Myr) before the epoch of observation, consistently with the scenario depicted by the AFM. They have assembled 
over their past 150-300 Myr, with short and repeated episodes of intense SF activity. 
Their progenitors are expected to be on average  faint in the rest-frame UV, even without invoking an extremely dust obscured phase. 
\item[$\bullet$] Galaxies with high \ml~at $z\sim7-8$ have assembled the bulk of their mass before $z\sim20$, i.e. 500–800 Myr prior to observation, indicating the presence of already-evolved stellar populations in the first few hundred megayears of our Universe.  They do not descend from the highest-redshift candidates known (e.g. $z\sim14$), but  their progenitors may be within reach of current observational facilities, with the majority expected to have a UV magnitude at $z>20$ brighter than $-20$ or $-17$, depending on the assumptions.
\item[$\bullet$]  The properties of quiescent galaxies depend on their stellar mass. 
The most massive systems have not undergone significant SF activity since $\sim 500$ Myr, 
while lower-mass systems formed the bulk of theirs stars more recently (over the past $\sim 300$ Myr), and are likely experiencing lulling phases in their SFHs. 
\end{itemize}

Synergy between long wavelength observations (with MIRI and ALMA) and large spectroscopic surveys is critical to validate the presence of significant dust in high-redshift systems. Furthermore, this strategy will help discriminate among the models proposed to explain the excess of UV-bright galaxies.

\begin{acknowledgements}
We acknowledge support from: 
INAF RF2024 Large Grant "UNDUST: UNveiling the Dawn of the Universe with JWST"; 
INAF RF2022 Mini Grant “The evolution of passive galaxies through cosmic time”; 
INAF RF2024 GO Grant ”Revealing the nature of bright galaxies at cosmic dawn with deep JWST spectroscopy”; 
INAF RF2022 Large Grant “Extragalactic Surveys with JWST”;  
PRIN 2022 MUR project 2022CB3PJ3 "First Light And Galaxy Assembly" (FLAGS) funded by the European Union NextGenerationEU; 
PRIN-MIUR 2020SKSTHZ grant; 
INAF Grants "The Big-Data era of cluster lensing'' and ``Probing Dark Matter and Galaxy Formation in Galaxy Clusters through Strong Gravitational Lensing".
PB acknowledges financial support 
from the Italian Space Agency (ASI) through contract ``Euclid - Phase E''. 
MB acknowledges support by the ANID BASAL project FB210003. This work was supported by the French government through the France 2030 investment plan managed by the National Research Agency (ANR), as part of the Initiative of Excellence of Université Côte d’Azur under reference No. ANR-15-IDEX-01. 
BS acknowledges support through an Erasmus Mundus Joint Master (EMJM) scholarship funded by the European Union in the framework of the Erasmus+, Erasmus Mundus Joint Master in Astrophysics and Space Science – MASS. Views and opinions expressed are however those of the author(s) only and do not necessarily reflect those of the European Union or granting authority European Education and Culture Executive Agency (EACEA). Neither the European Union nor the granting authority can be held responsible for them. 
This work is based  on observations made with the NASA/ESA/CSA James Webb Space Telescope. The data were obtained from the Mikulski Archive for Space Telescopes at the Space Telescope Science Institute, which is operated by the Association of Universities for Research in Astronomy, Inc., under NASA contract NAS 5-03127 for JWST. These observations are associated with programs \#1324, 1345, 1180, 1210, 1837, 2079, 2561, 2756, 3703, 3990. 
Some of the data products presented herein were retrieved from the Dawn JWST Archive (DJA). DJA is an initiative of the Cosmic Dawn Center (DAWN), which is funded by the Danish National Research Foundation under grant DNRF140. 
This work has made use of the flathub data hub at the Flatiron Institute, which is supported by
the Simons Foundation.
\end{acknowledgements}

%
%
\bibliographystyle{aa}
\bibliography{biblio}

\begin{appendix}

\section{Dependence on SED fitting assumptions} \label{app:fitsystematics}

It is commonly known that the SED fitting assumptions may affect the inferred physical parameters \citep[e.g.][]{pacifici23,wang24,harvey25}. This is particularly true for the choice of the SFH \citep[e.g.][]{ciesla17,leja19,lower20,haskell24}.  
In order to verify that our main results are robust and independent of SED fitting systematics, we repeated the analysis 
with different assumptions.

First of all, we tested the effect of our parameter grid choice by 
{\it i)} decreasing the minimum age to 50 Myr, {\it ii)} increasing the maximum \ebv~on the lines to 2, and {\it iii)} assuming the stellar metallicity to be 20\%, 1$\times$ and 2$\times$ Solar. None of these changes affect 
the results. The only appreciable difference is 
obtained in case  {\it iii)}, where, due to the lack of the low-metallicity templates, the sSFR shifts  upwards by 0.15 dex on average at $z>10$. This, however, does not change our conclusions. 

We then tested different assumptions 
for the SFH. In addition to our reference model {\it a)}, i.e. non-parametric model, with linearly spaced time binning, $t_0=20$ Myr and flat prior, we considered the following cases:

{\it b)} same assumptions as {\it a)}, but fixing the latest bin duration $t_0$ to 10 Myr;

{\it c)} same  as {\it a)}, but with  log spaced time binning;

{\it d)} non-parametric model similar to {\it a)}, but adopting a bursty-continuity prior \citep[e.g.][]{leja19}, which avoids sharp variations in the SFH and is thought to be less prone to the outshining effect \citep{wang25};

{\it e)}  same  as {\it d)}, but with  log spaced time binning \citep{ciesla23};

{\it f)}  a parametric delayed SFH (see Eq. 4 in \citealt{boquien19}) characterised by $\tau_{main}$= 10, 100, 500, 1000 Myr, plus a recent burst having $\tau_{burst}$= 10, 25, 50 Myr and age$_{burst}$ = 5, 10, 50 Myr, with the mass fraction produced in the recent burst ranging from 0 to 60\% (7 linearly spaced steps); in this case, we consider the SFR averaged over the last 10 Myr;

{\it g)}  a parametric delayed SFH as in {\it f)}, but the flexibility is accounted for by a recent period of constant SFR \citep{ciesla17}, lasting for 10, 50, or 100 Myr, which is a factor of $r_{SFR}$ higher or lower than the SFR immediately before the constant phase, with $r_{SFR}$=0.1, 1, 5, 10, 50, 100;  again,  we consider the SFR averaged over the last 10 Myr.

\begin{figure}[t!]
    \centering
\includegraphics[width=0.5\textwidth]{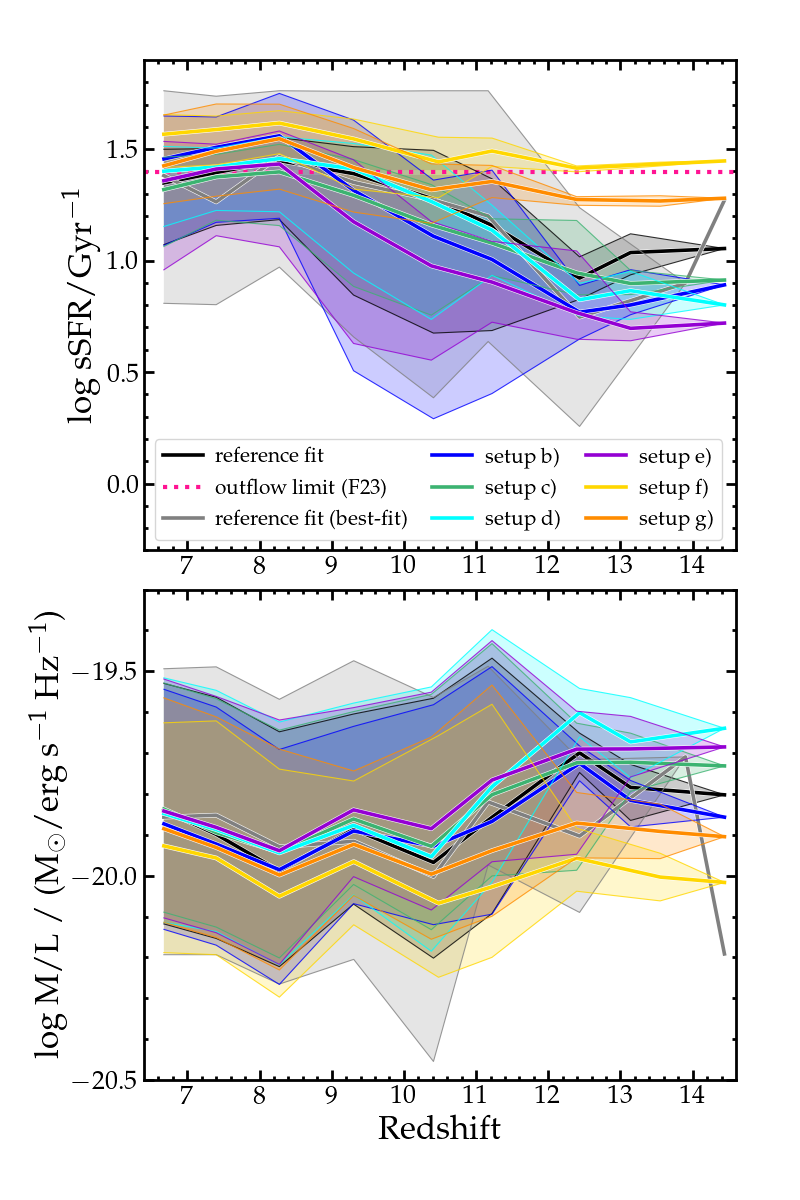}
\caption{Median (thick solid lines) and 16${th}$--84${th}$  percentile range (shaded areas) of the  specific SFR ({\it top} panel) and \ml~({\it bottom} panel) vs redshift for $8\leq$ \lgm~$<9$ galaxies according to different SED fitting assumptions and methodologies, as detailed in the legend.
}
\label{fig:ssfr_median}
\end{figure}

In the top panel of Fig.~\ref{fig:ssfr_median} we show the evolution of the median sSFR 
for $10^8-10^9M_\odot$ galaxies under the different assumptions described above. 
Overall, non-parametric fits have a higher level of scatter, as expected since they aim at reproducing a broader variety of sources.
Moreover, larger errors are expected when using non-parametric models compared to parametric ones, as the latter a priori rule out a significant fraction of possibilities \citep{leja19,haskell24}. Additionally, non-parametric fits show on average lower sSFR, with the various setups spanning 0.7 dex range for the highest-redshift sources. This is likely a consequence of mitigation of the outshining effect, responsible for underestimating stellar masses \citep{papovich01,leja22}.  
While  
investigating in detail the origin of the differences among fits is beyond the scope of the present work, all of them consistently predict a flattening or decreasing trend at $z\gtrsim 8-9$, 
corroborating the robustness of our result.

A similar conclusion can be drawn for the \ml~evolution, shown in the lower panel.  The difference among the various setups is lower in this case (0.3 dex at the highest redshifts), which is again not unexpected, as the rest-frame UV luminosity is a much more stable parameter than the SFR, which is more affected by different SFH assumptions and dust attenuation. 

We also tested that our results are robust against the methodology: if the best-fit solution is adopted instead of the median of the probability distribution function, results are overall unchanged, despite a larger scatter in the relations, as can be seen in Fig.~\ref{fig:ssfr_median}. The adoption of the median value filters out outliers caused by extreme models. 

While the SFH parameterisation is likely the variable mostly affecting the SED fitting results, we also tested the effect of modifying other assumptions, one at the time, keeping the choice of the SFH fixed. 
We first allowed a deviation from the Calzetti attenuation law, modified by means of a power law \citep{salim18} whose slope is allowed to vary between -0.3 and 0.3, in 5 equally sampled steps. We then assumed a \cite{cf00} attenuation law, letting attenuation in the ISM $A_V$ to vary between 0 and 0.8 (7 steps) and the attenuation in the birth clouds $A_{BC}$  regulated by the parameter $\mu=A_V/(A_V+A_{BC})$, ranging from 0.2 to 0.5 in steps of 0.1. The power law slope of the attenuation in the ISM and in the birth clouds is set to -0.7 and -1.3, respectively.
As a third step, we considered the BPASS v2.2 stellar population models \citep{eldridge17,stanway18}, which incorporate the effects of binary stellar evolution. 
Finally, to account for systematics associated with the nebular prescription, we tested the extreme (and unrealistic) case of ignoring the nebular contribution. 
While masses and SFRs are affected to a different level in each of these tests, with variations from a few percent to $\sim$60\% at most, and at different redshifts, the overall effect on the sSFR is very minor, and produces similar trends to those shown in Fig.~\ref{fig:ssfr_median}. Similarly, the flat observed evolution of the \ml~is robust against changing the SED fitting assumptions.  
In all these fits we kept the IMF fixed. We note that changing the IMF would result in a constant scaling in the masses (hence \ml) and SFRs, both in the same direction, hence only in minor shifts in the sSFR, once again not affecting the main conclusions of the paper.

\section{Accuracy of the inferred physical properties}\label{app:accuracy}

To verify the accuracy of the inferred SFR, we first compared our estimates to those inferred from emission lines for the spectroscopic 
sample of \cite{dottorini25}, made of NIRSpec sources extracted from the DJA archive.  
The common sample counts 680 sources. 

We calculated the spectroscopic SFR following \cite{calabro24}. 
Briefly, we computed the gas phase attenuation either from the H$\alpha$/H$\beta$ Balmer decrement or from the SED fitting ($A_{\rm V}=4.05 \times$\ebv$_{\rm line}$)
in case only one line is available, 
and we calculated the dust-corrected H$\alpha$ luminosity. 
For galaxies lacking H$\alpha$, we estimated its intrinsic luminosity from the dust-corrected H$\beta$, rescaled assuming an intrinsic H$\alpha$/H$\beta$ ratio of 2.86 \citep{osterbrock89}.  
 We then converted  the H$\alpha$ luminosity into a SFR estimate adopting the calibration of \cite{reddy22},  
 more appropriate for sub-Solar metallicity galaxies, typical at $z>7$.

\begin{figure}[t!]
    \centering
\includegraphics[width=0.52\textwidth]
{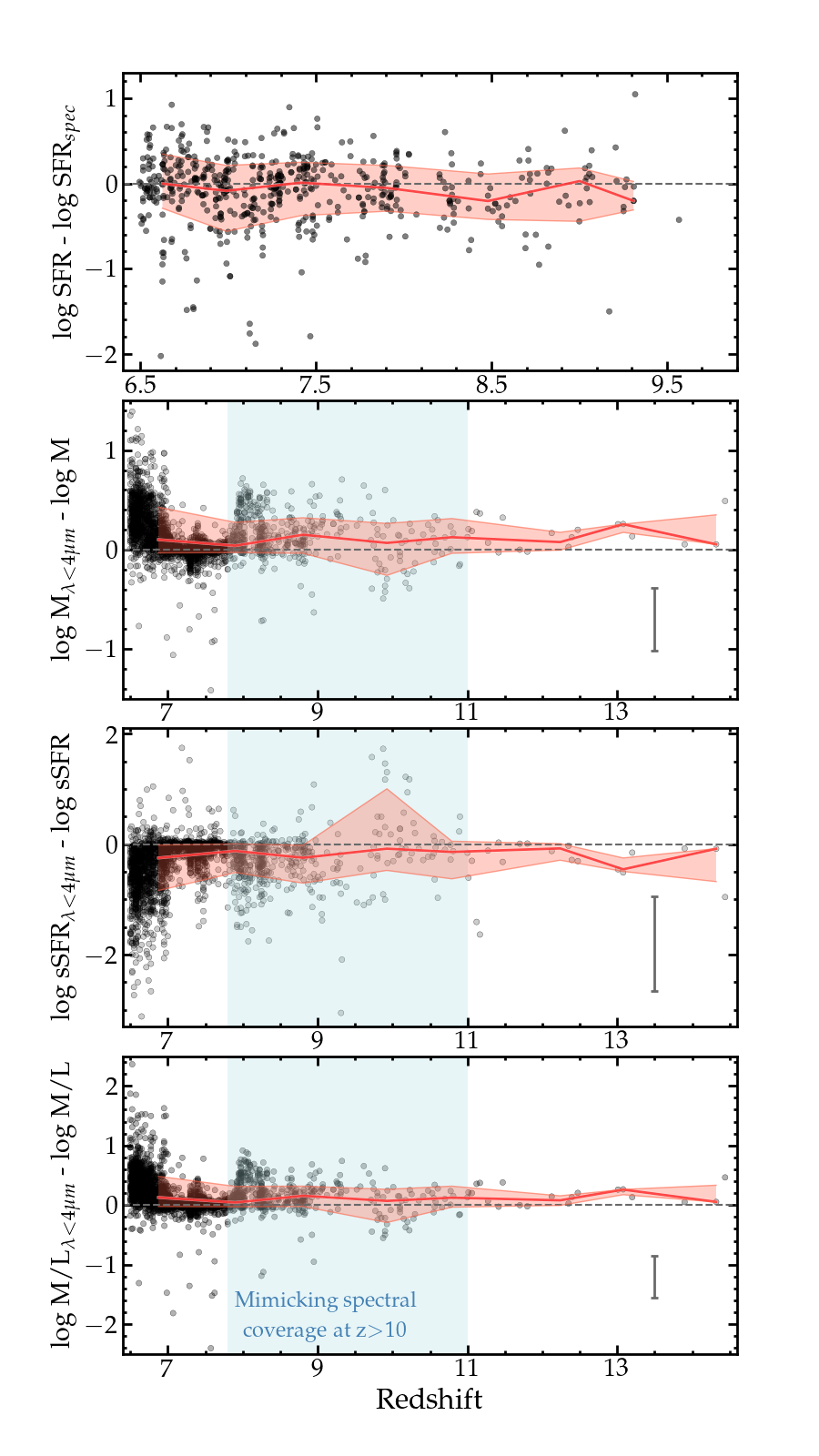}
\caption{{\it Upper} panel: Comparison of the SFR with spectroscopic values calculated from the sample of \cite{dottorini25}. 
Other panels: Comparison of the stellar mass (second row), sSFR (third row), and \ml~({\it bottom} row)  with that inferred ignoring the two reddest filters, to mimic  in the $z\sim 8-11$ range, shaded in light blue, the absence of rest-frame optical constraints affecting $z>10$ galaxies. The error bar shows the median  error 
obtained by symmetrising and propagating the errors on the physical properties. 
In all panels, 
the red line and shaded area show the median and the 16${th}$--84${th}$ percentiles range of all galaxies in bins of redshift. }
\label{fig:check}
\end{figure}

The comparison between SFRs calculated from SED fitting and those derived from spectroscopy is shown in the upper panel of Fig.~\ref{fig:check}. While individual galaxies can differ by  more than a factor of 10, the median offset in the logarithmic space is very close to zero, with a maximum deviation of $\sim$0.2 
dex in the $8.25-8.75$ redshift bin.

Once the reliability of SED fitting as an SFR tracer has been verified, another potential issue affecting our results lies in the fact that the available photometry samples the optical rest-frame only up to $z\sim10$, where the F444W filter encompasses the 4000\AA~break. 
At higher redshift, the rest-frame optical SED is unconstrained.  
To test whether this introduces any systematics in the estimated stellar masses and sSFR, we repeated the fit for the full sample by ignoring any filter redder than the F356W, i.e. the F410M and F444W filters. This setup  mimics the absence of optical constraints suffered by $z\sim 10-14$ galaxies in the $z\sim8-11$ redshift range.  Specifically, the F444W filter extends up to rest-frame $\sim 4500$ (3300) \AA~at $z\sim 10$ (14), while the F356W filter probes up to  $\sim 4400$ (3300) \AA~at $z\sim 8$ (11).

Although mass uncertainties increase, with relative uncertainties being on average $\sim$15\% higher and reaching $\sim$50\% at $z\sim10$, the mean stellar masses are only mildly affected. 
The ratio of the stellar mass inferred without rest-frame optical bands to those calculated exploiting the full photometry is always very close to 
 unity, as shown by the second row panel of Fig.~\ref{fig:check}.  
 In the range of interest ($z\sim8-11$, highlighted in light blue in the figure), the median offset is $\sim$0.1 dex, well within the typical uncertainties on stellar masses. 
We can conclude that for the vast majority of galaxies the lack of the reddest filters does not affect the stellar mass estimates significantly, qualitatively in agreement with the results of \cite{helton25b}, who tested the effect of MIRI photometry on the inferred physical parameters of $z\sim8$ galaxies (but see also divergent results from \citealt{wangTao25} and \citealt{cochrane25}). 
We note however that  
 the distribution is highly asymmetric at specific redshifts, with a subset of sources in the tail of the distribution showing 
differences as large as a factor of $5-10$. 
This occurs 
when the two reddest filters encompass the H$\beta$ and [OIII] lines (i.e. at $z\sim 8$), 
 and the full set of filters 
 allows the SED modelling to properly constrain nebular emission in these galaxies. 
We verified that the sources whose stellar masses are substantially larger when the two reddest filters are excluded show high equivalent widths, and therefore lower stellar masses, when all bands are included in the fit.  
We note that this issue affects only a minority of sources, with only 2\% of the total sample showing a difference larger than a factor of 5. 
 
The third and bottom row panels of Fig.~\ref{fig:check} show a similar comparison for the sSFR and the mass-to-light ratio, respectively. An offset of $\sim 0.1$ dex in the opposite direction compared to the one for stellar mass is observed for the sSFR, suggesting that SFRs are not significantly affected by the lack of optical constraints. 
A similar conclusion can be drawn for the \ml, with an overall trend following the one shown by stellar masses, as expected. 

For all three properties (M, sSFR, and \ml), the only significant deviation is observed at $z>12$, although still within the typical uncertainties. 
However, deriving the properties of $z\sim12$ galaxies without the two reddest NIRCam filters is beyond our purposes, as it would  simulate our ability to properly model galaxies outside the range of interest.  At this redshift the F356W filter samples the rest-frame region shortwards of 2700\AA, equivalently as the F444W filter does at $z\sim15$. 

In conclusion, the physical properties of galaxies in the critical $z\sim8-11$ interval - where the spectral coverage without the two reddest filters matches that of our $z\sim10-14$ sample - are not affected by the lack of optical constraints. Only a small fraction of sources (a few percent) is affected, which does not impact the overall distribution.

\section{Simulation to assess the selection window and completeness} \label{app:completeness}

As discussed in Sect.~\ref{sec:completeness}, given the complex selection criteria, we assessed the selection window and completeness of our sample through simulations. The same simulations are also used to evaluate our ability to retrieve galaxy properties.

We first considered the simulated catalogues from the Santa Cruz semi-analytical model  \citep[SC-SAM hereafter,][]{somerville21,yung22}.  
 These mock catalogues are built by combining dissipationless N-body simulations with the Santa Cruz semi-analytic model \citep{somerville15,yung19a}, with the aim of aiding the interpretation of observational data, and have been shown to fairly reproduce observed galaxy properties at least up to $z\sim 10$ \citep{yung19a,yung19b,somerville21}. 
In particular,  we used the SC-SAM JWST deep-field model, made of eight ultra-deep lightcones (132 arcmin$^2$) probing galaxies up to $z\sim12$ at depths comparable to the deepest JWST surveys, i.e. resolving 
down to  $\sim$34 mag in the  F200W band or rest-frame \muv $\sim$-14.  
We queried the FlatHUB\footnote{\url{https://flathub.flatironinstitute.org}} archive and selected galaxies in the 6.5--15 redshift range with UV absolute magnitude (uncorrected for dust reddening) between $-16.5$ and $-23.5$ and stellar mass larger than $10^7M_\odot$. In addition to galaxy properties, the SC-SAM provides  the expected photometry in the same bands used in this work with the exception of the HST WFC3 F140W band. We perturbed the predicted photometry according to the properties of the JADES-GS field, showcasing a compromise in terms of depth and area of the full dataset. We analysed the mock photometry in the very same way as observed galaxies: we calculated the photometric redshifts and associated $P(z)$, inferred the physical properties, and applied the selections described in Sect.~\ref{sec:sample}. 

Most model galaxies are fainter than the observed ones, and the S/N$>$5 cut in the F444W band is responsible for reducing the sample to 35\%. 
The $z>6.5$ threshold, the photometric cuts and the criteria based on the shape of the $P(z)$ further reduce the sample to 31, 28 and 22\%, 
respectively. 
Overall, from an initial sample of 54122 
mock galaxies we ended up with 11991 
passing all selection criteria.

\begin{figure*}[t!] 
    \centering
\includegraphics[width=\textwidth]{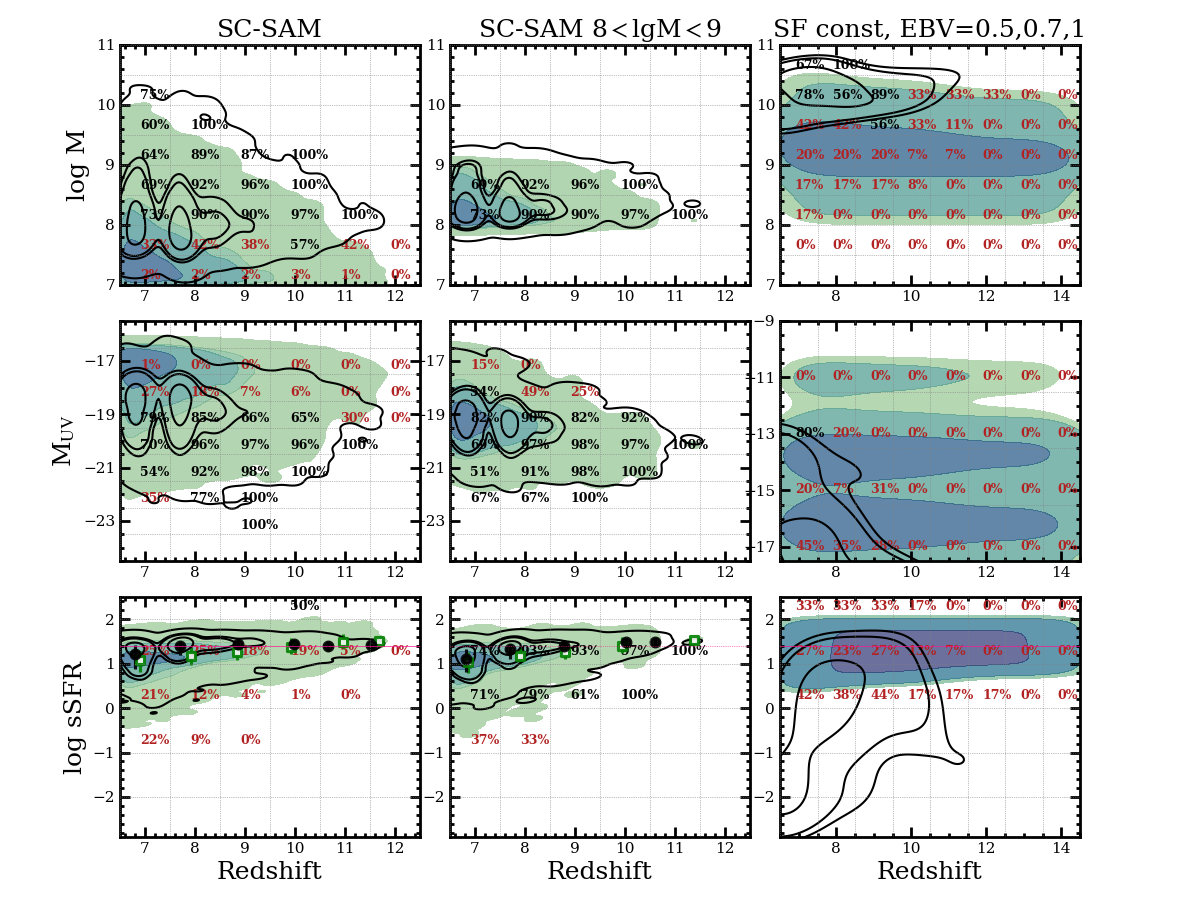}
\caption{The coloured regions  show the density 
of simulated galaxies (10, 40, 80, and 100\% )
at different redshifts and stellar masses ({\it top} panel), UV luminosities ({\it middle}) and sSFR ({\it bottom}). The {\it left} and {\it central} panels display galaxies from the SC-SAM with \lgm~$>7$ and limited to the $8\leq$ \lgm~$<9$ mass range, respectively, while the {\it right} panels represent simulated galaxies with constant SFR and heavily dust extincted. We note that the $y$-range in the central right panel is different, as such simulated galaxies are extremely faint at UV magnitudes.  In each bin of the grid we report the fraction of sources recovered by our selection; the red labels indicate a completeness lower than 50\%.  The black open contours show the distribution 
of the inferred properties for the mock galaxies. In the bottom right and central panels, the open green and solid black symbols show the median evolution of the sSFR considering the input  and the inferred properties, 
respectively. }
\label{fig:simu}
\end{figure*}

To evaluate the completeness of our selection, we considered bins of stellar masses, sSFR and UV luminosities at different redshifts. 
For each grid bin, we computed the ratio of the number of sources with the relevant inferred properties that pass the selection to the total number of input galaxies in the bin with the corresponding input properties.
We report the results in Fig.~\ref{fig:simu} (left panels).

As expected, the lowest completeness is found for low-mass ($<$$10^8 M_\odot$) and faint ($M_{UV}$>$-18$) galaxies, with less than 20-40\% of galaxies recovered by the selection. Redshifts close to the lowest cut also suffer from a lower than average completeness level ($\sim 60-70\%$), due to galaxies being rejected as soon as they scatter below the threshold. The rest of the sample is highly complete in stellar mass and UV luminosity, typically $>70-90\%$. The overall completeness in sSFR is quite low, but it is higher than 80-90\% when limiting to galaxies in the mass range typical of observed sources, i.e. $8\leq$ \lgm~$<9$ (central panels of Fig.~\ref{fig:simu}). 

The distribution of stellar mass, UV luminosity and sSFR are well reproduced by our selection fitting procedure, especially when limiting to the typical mass range of our observed galaxies, including the increasing trend of the sSFR with redshift (both the input and the output sSFR increase by $\sim$0.4 dex from $z\sim6.5$ to $z\sim 11$, bottom central panel of Fig.~\ref{fig:simu}). This implies that the observed lack of evolution in the sSFR is not an artefact of our SED fitting technique, nor of our selection provided that the simulation is representative of all relevant galaxy classes. 

The SC-SAM misses heavily dust-obscured objects and does not include dust at $z\gtrsim10$ at all. 
To overcome this, we complemented it with galaxies simulated from stellar population synthesis models. We  simulated galaxies with constant SFR, normalised to 10 and 100 $M_\odot/yr$, 10\% Solar metallicity, age ranging from 20 Myr to the age of the Universe at the relevant redshift and \ebv~equal to  0.5, 0.7 or 1. 
Again, we perturbed their photometry and treated these mock galaxies in the very same way as the observed sample. 
As shown in the right panels of Fig.~\ref{fig:simu}, not unexpectedly, our selection is unable to include these galaxies, in particular at $z>10$. Only  56\% 
of the mock galaxies  have the required significancy ($>5\sigma$) in the F444W band.  The other steps of the selection, i.e. the $z>6.5$ threshold, photometric cuts, and $P(z)$ constraints, progressively reduce the sample to 
41, 33, 15\%, 
respectively, of its initial size. 
Only a few massive (\lgm$\gtrsim$9.5) galaxies satisfy the selection criteria 
and the physical   properties of the overall dusty population are poorly reproduced.  
The well-known degeneracy between red-dusty sources and red-passive ones becomes  evident from the extremely low sSFR value recovered for these objects (bottom right panel). We note, however, that the 
observed quenched galaxies with sSFR $<$ 0.1/Gyr 
cannot be explained by this effect. The expected \ml~of simulated dusty galaxies erroneously fitted with low sSFR are 
extremely high, as a consequence of their extremely faint UV luminosity, whereas we do not recover similar \ml~in our fits.

When only the $z>10$ subsample is considered, the SED modelling lacks  optical rest-frame constraints and relies on the rest-frame UV emission, most sensitive to dust obscuration. The fraction of $z>10$ simulated galaxies passing this magnitude cut is only  34\%. 
While the first two of the subsequent criteria reduce the sample
to 
 20 and 16\% of the parent sample, 
 only $<$3\%
of the sources passes the $P(z)$ selection, implying that their photometric redshift is  almost never reliable enough to be considered in the final sample.

To conclude, our selection is significantly incomplete towards galaxies with a high level of dust obscuration, and it  almost completely misses them 
at $z>10$, where their existence 
has yet to be confirmed. 
Moreover, it suffers from severe incompleteness at stellar mass lower than $10^8 M_\odot$ and $M_{UV}>-18$. Our sample is instead reasonably complete at stellar masses between $10^8$ and $10^9 M_\odot$, i.e. where the observed mass distribution peaks (see Fig.~\ref{fig:Mdist}).

\section{Best fit of the mass-luminosity relations}\label{app:mlbestfit}

We used linear least-squares regression to examine the relationship between
the stellar mass and the rest-frame UV luminosity in different redshift bins. We parameterised the data as  $\log M=a \cdot (M_{UV}+19.5)+b$ and report the best-fit parameters in 
Table~\ref{tab:masslum}.

\begin{table*}
    \centering
    \begin{tabular}{cccc}
     \noalign{\smallskip}\hline\noalign{\smallskip}
        Redshift bin & $a$ & $b$ & Pearson coefficient \\
     \noalign{\smallskip}\hline\noalign{\smallskip}
        $6.5 \leq z<7.5$ & -0.37 $\pm$ 0.01 & 8.62 $\pm$ 0.01 & -0.68 \\
        $7.5 \leq z<9$ & -0.28 $\pm$ 0.02 & 8.53 $\pm$ 0.02 & -0.52 \\
        $9 \leq z<14$ &  -0.35 $\pm$ 0.04  & 8.65 $\pm$ 0.04 & -0.70 \\
         \noalign{\smallskip}\hline\noalign{\smallskip}
    \end{tabular}
    \caption{Best-fit parameters of the mass-luminosity relations, expressed as $\log M=a \cdot (M_{UV}+19.5)+b$. }
    \label{tab:masslum}
\end{table*}

\section{Assembly history of the galaxies with low SF activity}\label{app:tracks_lowssfr}

\begin{figure}[t!]
    \centering
\includegraphics[width=9cm]{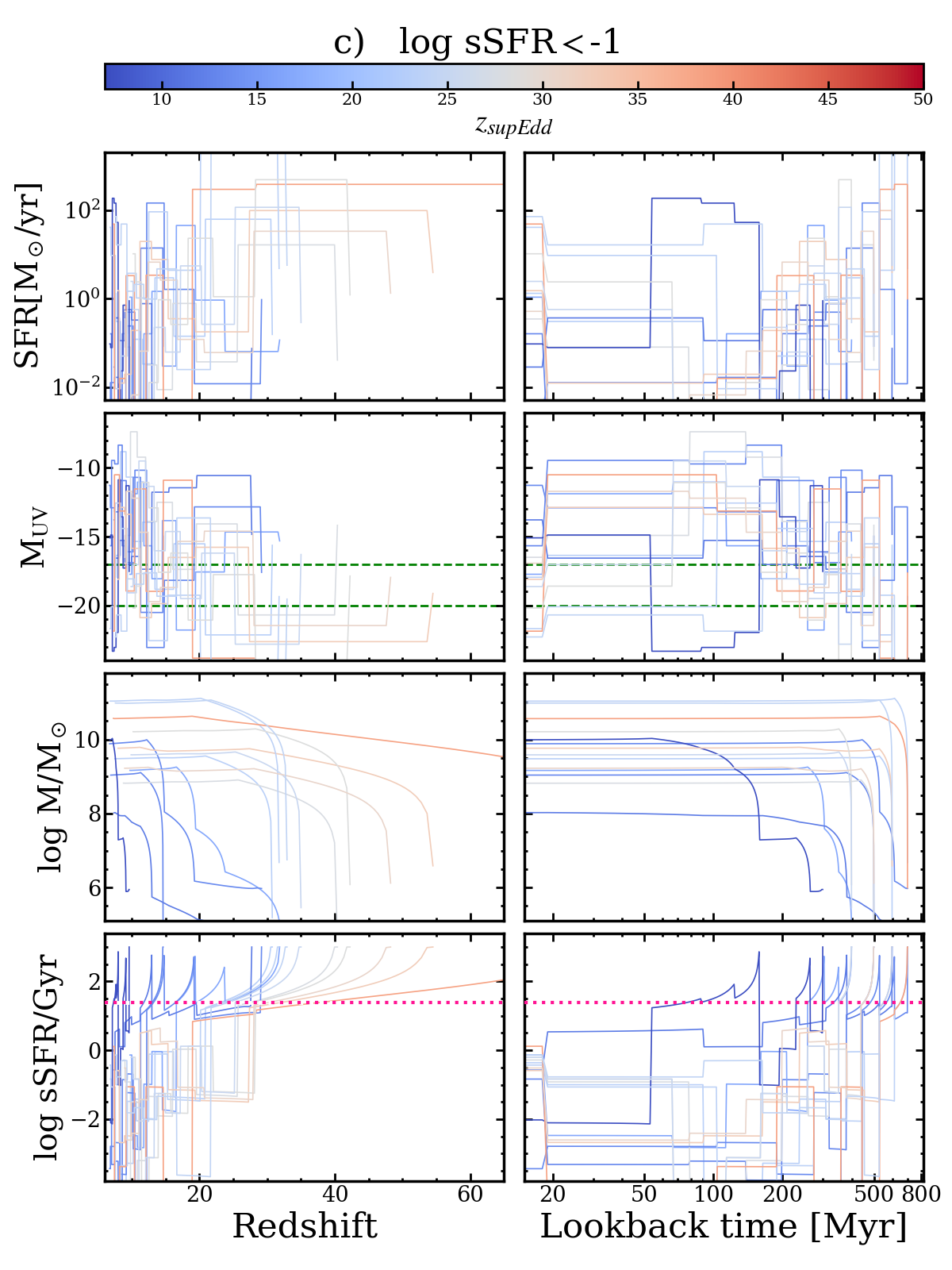}
\caption{Same as Fig.~\ref{fig:tracks}, but for galaxies with  sSFR lower than 0.1/Gyr.}
\label{fig:tracks_lowssfr}
\end{figure}

We show in Fig.~\ref{fig:tracks_lowssfr} the assembly histories of galaxies with sSFR lower than 0.1/Gyr, discussed in Sect.~\ref{sec:passive} and \ref{sec:progenitors}.

\end{appendix}

\end{document}